\documentclass[useAMS,usenatbib,fleqn]{mn2e}

\usepackage{times}

\usepackage{amsmath}
\usepackage{amssymb}
\usepackage{upgreek}
\usepackage{graphicx}
\usepackage{natbib}

\renewcommand{\vec}[1]{\ensuremath{\bmath{#1}}}   %Vector quantity in bold
\title[Radiation Driven Flares]{Maser Flares Driven by Variations in Pumping and Background Radiation}
\author[M. D. Gray, S. Etoka, A. Travis and B. Pimpanuwat]
{M. D. Gray$^{1,2}$, S.Etoka$^{1}$, A.Travis$^{1}$ and B. Pimpanuwat$^{1}$\\
$^{1}$Jodrell Bank Centre for Astrophysics, Department of Physics and Astronomy, University of Manchester,
M13 9PL, UK\\
$^{2}$National Astronomical Research Institute of Thailand, 260 Moo 4, T. Donkaew, A. Maerim, Chiangmai 50180, Thailand.}
\begin{document}

\date{}

\pagerange{\pageref{firstpage}--\pageref{lastpage}} \pubyear{2018}

\maketitle

\label{firstpage}

\begin{abstract}
We simulate maser flares by varying either the pump rate or the background level of radiation
in a 3D model of a maser cloud. We investigate the effect of different cloud shapes, saturation
levels and viewpoints. Results are considered for clouds with both uniform and internally variable
unsaturated inversion. Pumping and background 
variations are represented by several different driving functions,
some of which are light curves drawn from observations.
We summarise the pumping variability
results in terms of three observable parameters, the maximum flux density achieved, a variability
index and duty cycle. We demonstrate typical ranges of the flux density that may result from
viewing an aspherical object from random viewpoints. The best object for a flare is a
prolate cloud, viewed close to its long axis and driven from unsaturated conditions to
at least modest saturation. Results for variation of the background level are qualitatively different
from the variable pumping results in that they tend to produce short intervals of low
flux density under conditions of moderate saturation and sufficient variability
to be consistent with strong flaring. Variable background models typically
have a significantly higher duty cycle than those with variable pumping.
\end{abstract}


\section{Introduction}
\label{intro}

This paper is the third in a series that investigates several plausible mechanisms
for maser flares. \defcitealias{2018MNRAS.477.2628G}{Paper~1}
In \citet{2018MNRAS.477.2628G} \citepalias{2018MNRAS.477.2628G}, we investigated variability
produced by the rotation of a quasi-spherical maser cloud. That work was extended to
\defcitealias{2019MNRAS.486.4216G}{Paper~2}
the rotation of approximately spheroidal oblate and prolate clouds in 
\citet{2019MNRAS.486.4216G} \citepalias{2019MNRAS.486.4216G}. The present work introduces the 
mechanisms of time-dependent variation in the pumping scheme and time-dependent
variation in the background radiation amplified by the maser cloud. We do
not rule out cloud overlap in the line of sight as a flare mechanism, but defer analysis
of the overlap mechanism to a later paper in this series. 

We define a variability index for the purposes of this work as $F_{pk}/F_{qui}$,
where $F_{pk}$ is the line centre flux density at the flare maximum and $F_{qui}$ is 
the corresponding flux density under quiescent conditions. If the light curve has no
flat quiescent state, $F_{qui}$ defaults to the minimum flux density.
The oblate and prolate clouds studied in \citetalias{2019MNRAS.486.4216G} can produce flares with variability
indices of thousands in optimal viewing planes, but more typically of order tens for a
randomly placed observer. Rotation of the pseudo-spherical object in \citetalias{2018MNRAS.477.2628G}
produces variability index values of $\sim$3.
Stability considerations strongly suggest that rotation
is most unlikely to produce periodic flares, but both the flare mechanisms considered
here would naturally follow any periodicity in a source of pumping (or background)
radiation.

\subsection{Observational Background}
\label{obsback}

At least some flaring variability in astrophysical masers is almost certainly driven
by changes in the flux of external pumping radiation incident on a maser zone. The main
evidence for this statement comes from correlation between maser flaring events and
changes to the flux of infra-red (IR) radiation, from sources close to the maser, that
incorporate wavelengths important in the pumping scheme of the relevant molecule. For
example, power in the 6.7-GHz maser transition is highly correlated with dust emission
at 870\,$\umu$m \citep{2015MNRAS.446.3461U}.
We note that in multi-level systems, IR wavelengths corresponding to individial molecular
transitions may cause either an increase or a decrease in maser brightness, depending on
the details of the pumping scheme. Variation of the IR pumping radiation has been linked,
in at least some massive protostellar sources, to enhanced accretion events
\citep{2019BAAS...51c..13H,2017ApJ...837L..29H}. Most of the work described here relates to masers in star-forming
regions, but we note that flaring variability in H$_2$O \citep{2018IAUS..336..393B} and
OH \citep{1997A&A...321..877E,2017MNRAS.468.1703E} masers 
has also been observed
in evolved stars, in addition to lower amplitude variations
linked to the pulsation period, and associated stellar continuum variations, in these
sources.

Specific examples of IR/maser flare correlations include the 22-GHz H$_2$O 
and 6.7-GHz methanol masers associated with the ucH{\sc ii} region G025.65+1.05
\citep{2019RAA....19...38S,2017ATel10842....1S}. The H$_2$O maser variability in this object may be correlated
with IR K-band variability in a source 
approximately 1500\,AU from the maser position \citep{2019RAA....19...38S}. To be precise,
the H$_2$O emission appears to exhibit an anti-correlation, with bright 22-GHz episodes corresponding to 
times of low K-band (most observations centered on 2.20\,$\umu$m wavelength) flux density. This 
is to be expected for typical 22-GHz masers
in star-forming regions, where computational models, for example \citet{2016MNRAS.456..374G},
show that strong IR emission
from dust tends to destroy the 22-GHz inversion, probably by inhibiting the loss
of population from the lower maser level \citep{1981SvA....25..373S,2012IAUS..287...13S}.
The 6.7-GHz methanol masers in this source have been described as showing a moderate flux
rise at the time of a 22-GHz H$_2$O maser flare
\citep{2017ATel10757....1S}.
A radiative mechanism is not the only explanation for H$_2$O flaring in G025.65+1.05. VLBI observations
identify the flaring object with a single compact spot of maser emission \citep{2020MNRAS.491.4069B}. An image
from spectral channels in the blue-shifted wing of this spot shows a double structure, leading
to an interpretation as a possible line-of-sight overlap of two masing objects. Preliminary
 modelling results for spatial overlap \citep{2018evn..confE..73G} indicate that this mechanism 
can produce flares of significantly
higher variability index than is typical of radiative pumping mechanisms, and the overlap interpretation
therefore becomes more likely if the source is at the further of the two possible distances (2.7 and 12.5\,kpc) mentioned
 in \citet{2020MNRAS.491.4069B}. Another possible example
of this overlap mechanism is a very high brightness H$_2$O flare from Orion~KL
\citep{2005ApJ...634..459S}.
The 22-GHz maser emission of G025.65+1.05 has been monitored by a combination
of the Simeiz, Torun and Hartebeesthoek
telescopes since 2000 \citep{2019ARep...63...49V}. Observations were approximately monthly for early data,
but on an almost daily basis since a giant flare in 2017-18. During the
period 2000-2019, it has flared strongly (peak flux density $>$ 2000\,Jy) three times: the peak flux density of each flare shows an
increasing trend, whilst the flare duration and inter-flare interval have decreased. The
shortest, brightest and most recent flare (2017-18) showed a double-peaked light curve, with
rising sections well-approximated by exponentials with characteristic times of approximately
10\,d. The exponential shape, backed by evidence from polarization and spectral width, have
led \citet{2019ARep...63...49V} to conclude that the flaring maser is unsaturated. The
continuum flux at 870\,$\umu$m has generally been rising towards G025.65+1.05 since 2008,
but the data is very sparse compared to the maser observations, including an interval of eight
years with no data \citep{2019ARep...63...49V}.

Although correlations between IR radiation and maser flux density have been demonstrated
in some sources, observations of the IR radiation have rarely been carried out with
sufficiently high cadence to reconstruct accurately the IR light curve associated with
a maser flare, though this has been attempted in K-band observations of 
G025.65+1.05 and G107.298+5.63 \citep{2018msa..conf...81B}. This is just a consequence
of the inherent difficulty of observing maser flares: their association with high-mass
young stellar objects (YSOs) makes them rather rare and, apart from
the known periodic sources, not simply predictable. Perhaps the only case in which a high-cadence
(sample separations of hours to weeks) IR light curve has been recorded with the
same period (34.6\,d) as a flaring maser is that of the intermediate-mass YSO G107.298+5.63, observed
with NEOWISE \citep{2018IAUS..336...37S}. NEOWISE records in four IR bands, W1-W4, with
respective band-centre wavelengths of 3.4,4.6,12 and 22\,$\umu$m \citep{2015ApJ...814..117N}.
The IR light curve itself has a saw-tooth form, consisting of a steep rise ($\sim$5\,d) to maximum light,
followed by a relatively slow decay that takes approximately 30\,d. Possible models for this
source are discussed by \citet{2014PASJ...66...78F}. An accretion-pulsation model of a YSO \citep{2013ApJ...769L..20I} 
would require a much more luminous YSO for the observed period. However, a colliding-wind binary
model \citep{2009MNRAS.398..961V} was found to agree reasonably well with the maser data. This model has a rapid rise
and slow decay, and is therefore also consistent with the IR data from \citet{2018IAUS..336...37S}.
The 34.6\,d period is also shared by 22-GHz H$_2$O masers \citep{2016MNRAS.459L..56S} that
vary in anti-phase with the 6.7-GHz methanol masers discussed above. Their EVN observations were carried out near one
peak in the methanol maser emission, and thirteen distinct clouds were detected, in an approximately
linear distribution of 400\,mas (306\,AU) in extent, parallel to a CO outflow.

Another interesting periodic ($P=243.3$\,d) source is G9.62+0.20E, which has correlated flares in transitions
of OH and methanol \citep{2019MNRAS.485.4676G}. This was one of the original periodic methanol maser sources
monitored by \citet{2004MNRAS.355..553G}. G9.62+0.20E also contains flaring H$_2$O masers, but
these are aperiodic. The particular transitions are the 6.7 and 12.2\,GHz transitions of methanol
and the 1665 and 1667\,MHz transitions of OH. Comparison with VLBI data from \citet{2015ApJ...804L...2S}
shows that only a subset of the OH maser spots are involved in the periodic flaring. The shapes and
timings of the OH and methanol light curves are different, with the OH transitions exhibiting an
initial dip in flux density before rising to peak output. The OH dip appears to coincide with
the start of the 12.2-GHz methanol rise, but the OH peaks were delayed with respect to the
12.2-GHz methanol peak by 23\,d (1665\,MHz) and 13\,d (1667\,MHz) \citep{2019MNRAS.485.4676G}. These
delays are attributed by \cite{2019MNRAS.485.4676G} to a $\sim$1600\,AU position difference between
the OH and methanol maser regions, with the driving mechanism in this source being variation in
the free-free background radiation of the masers from a nearby H{\sc ii} region. The periodicity
was linked ultimately to colliding-wind binarity. Maser decay from the
flare peaks was fitted to a recombination model by \cite{2009MNRAS.398..961V}, but there
is currently no measured continuum light curve against which this theory can be tested. An alternative
IR driving mechanism, also ultimately dependent on binarity, was suggested by \citet{2015ApJ...804L...2S}.

Radiative flare mechanisms have also been suggested for recurrent, but non-periodic, maser
flares, such as the multi-species, multi-transition event in NGC 6334I \citep{2018MNRAS.478.1077M}, which
was found to be correlated with a sub-mm continuum outburst associated with a rapid rise in dust 
temperature. Binarity is again discussed by the authors as the likely driver, but in this source the orbit is
decaying, leading to progressively shorter inter-flare intervals if the flares are associated with
a dust heating event at each periastron. Another source with an IR-methanol maser correlation
is S255 NIRS3 (Caratti o Graratti, Cesaroni et al. \citeyear{2017MmSAI..88..773C};
\citealp{,2017A&A...600L...8M}). This source is believed to contain
a massive protostar of estimated mass 20\,M$_{\sun}$, that suffered an accretion outburst, visible
in the IR, beginning around June 2015. Radiation in the 20-30\,$\umu$m waveband is necessary to
pump the 6.7-GHz methanol maser \citep{1997A&A...324..211S}. Radiation in this wavelength
range appears at pre-outburst and burst levels in the spectral
energy distribution shown by Caratti o Garatti, Stecklum et al.
\citeyearpar{2017NatPh..13..276C}, though data at higher spatial resolution
were available only in the shorter wavelength IR at the K, K$_s$ and H-bands. A high-cadence
light-curve in the K$_s$ band has recently become available \citep{2019PASJ..tmp..129U}.
The source was monitored for 8.5\,yr with the Torun 32-m telescope, extended to approximately
23\,yr with archival data \citep{2018A&A...617A..80S}. Over this time, only modest (20-30\%)
variability had been noted in the maser flux density prior to the burst of 2015-16.
Interferometric (EVN
and JVLA) observations of the associated 6.7-GHz methanol masers \citep{2017A&A...600L...8M} show
that a new flaring cluster of masers formed approximately 1000\,AU from NIRS3 in sky projection, whilst
a pre-burst maser cluster closer to NIRS3 ($\sim$400AU) was destroyed. A minimum propagation speed
of 0.02$c$ eliminates any shock-wave mechanism for the flare, and requires a radiative link. In
common with other sources, the flaring masers occupied a new region in velocity space, whilst
other pre-burst spectral features are not affected during the flare. A comparison of single-dish,
JVLA and VLBI data by \citet{2018A&A...617A..80S} demonstrates that pre-burst and bursting
maser regions in S255 NIRS3 are substantially more extended than the VLBI spots of a few milliarcsec.
\citet{2017A&A...600L...8M} estimate the extent of the flaring maser region: 0.24\,arcsec, or 430\,AU.

If IR pumping radiation always destroys the inversion in a particular transition, as might
be supposed for the 22-GHz transition of H$_2$O if always collisionally pumped, but increases
the inversion in another, such as the 6.7-GHz transition of methanol, then it would be possible
to distinguish between variable pumping and variable background mechanisms in any source containing
these two transitions, if suitably related spatially. A variation in the IR pumping radiation would
generally lead to an anti-correlation between the two transitions, whilst a variable background would
cause a positive correlation. G107.298+5.63 is an example of the first source type
\citep{2016MNRAS.459L..56S}, whilst NGC 6334I \citep{2018MNRAS.478.1077M} is an example of the
second. However, it is not clear that sources can be distinguished 
unambiguously in this way because the 22-GHz
transition is part of a family of H$_2$O transitions that have both radiative and collisional
pumping schemes \citep{2016MNRAS.456..374G}. The radiative regime requires, in
addition to dust temperatures above 850\,K, number densities of about
$3\times 10^{10}$\,cm$^{-3}$, roughly an order of magnitude higher than the collisional regime, but
substantially lower kinetic temperatures of $<$300\,K (compared to typically 700-1200\,K in the
collisional regime).
Overall, observed anti-correlation between these transitions almost certainly implies a
variable IR pumping mechanism, but a positive correlation could arise from either mechanism.

It should also be noted that non-maser explanations exist for flaring molecular line
emission. Perhaps the most likely of these is the Dicke superradiance model, see for
example \citet{2017SciA....3E1858R}. This model relies on the difference between the
spontaneous emission characteristics of isolated atoms or molecules, and their behaviour
in bulk, where the probability of spontaneous emission depends on a density matrix that
represents the statistical properties of a large number of molecules packed into a
distance small compared to the wavelength of the emitted radiation.

\subsection{Timescales and Cloud Sizes}
\label{ss:timescales}

An important parameter for the modelling work described in this paper is the shortest
timescale on which flaring variability occurs (see Section~\ref{ss:tscale}) and its relation
to another important parameter, the linear size of the maser cloud. A useful recent survey
in this respect is \cite{2018ARep...62..584S} who monitored seven H$_2$O maser sources during
April-September 2017 with the 22-m Simeiz telescope at an approximately monthly cadence.
All these sources exhibited variability, and very large flux density changes were recorded
on the shortest spacings between observations (20-30\,d). Size estimates from related interferometric
observations in two of the sources returned maser spot sizes of $\sim$2\,AU in
G34.403+0.233 \citep{2011PASJ...63..513K} and 2.7\,AU in G43.8−0.13 from 2015 RadioAstron data.
These spot sizes can possibly be regarded as typical, but neither
can be directly identified with the flaring objects. In the case of the object with the known high-cadence
IR light curve, G107.298+5.63, a parallax distance of 750$\pm$27\,pc was calculated by
\citet{2008PASJ...60..961H} so 1\,mas in this source corresponds to a linear distance
of 0.75\,AU. Figures in \citet{2008PASJ...60..961H} that show individual H$_2$O maser spots are
therefore consistent with diameters of $\sim$2\,AU. No sizes were available for the methanol
masers in this source, but the periodic source, G9.62+0.20E, was imaged with the VLBA in
the 12.2-GHz methanol maser line by \citet{2005MNRAS.356..839G}. These 12.2-GHz observations
detected 16 components, correlated in position and Doppler velocity, with compact core sizes
of $\sim 2$\,mas. However, G9.62+0.20E is at 5.7\,kpc, so the linear sizes of methanol maser
clouds could be significantly larger (of order 10\,AU) than those supporting H$_2$O masers.
Analaysis of the detected components showed no abrupt motion changes, or appearance of new
components, again suggesting a radiative driving mechanism, rather than, for example, a shock
impact.

Measured VLBI spot sizes may give a poor estimate of the extent of the
physical object through which a maser amplifies unless the observation includes short baselines.
Estimates from \citet{2017A&A...600L...8M} for the typical sizes of 6.7-GHz methanol maser spots are 2-20\,AU for
compact cores, with haloes of weaker emission extending 10-300\,AU. The much larger extent (430\,AU)
of the flaring emission in S255 NIRS3 is probably not problematic, since this maser appeared to
be unsaturated, yielding a light-crossing time, for the maser radiation, of $\simeq$2.4\,d. Overall,
$<1$\,AU-scale clouds are probably typical for H$_2$O masers, based on clustering 
scales from \citet{2010ApJ...715..132U}, but sizes an order of 
magnitude larger are likely typical
for methanol (2-30\,AU), based on measurements in 13 sources \citep{2008PASJ...60...23S}
while larger-scale structures cannot be ruled out. In Cepheus~A, \citet{2010ApJ...715..132U} find
a clustering scale of 0.31$\pm$0.07\,AU for H$_2$O masers. In the same source, methanol 6.7-GHz features
from \citet{2011A&A...526A..38T} have a scale of $\sim$5\,milliarcsec (3.5\,AU) which is consistent
with the methanol clouds being systematically larger.

\subsection{Flare Mechanisms}
\label{ss:flarmech}

In this work, we are primarily concerned with the detailed response of a 
maser cloud to a known driving light curve for pumping or background radiation, under 
a range of possible levels of saturation, and the variation of the response with
the viewing direction of the observer. We are less concerned with the physical
mechanism, involving a binary system or otherwise, that produces the pumping
radiation, unless the theory is sufficiently developed to predict the functional
form of the light curve. The observational IR light curve from G107.298+5.63 \citep{2018IAUS..336...37S} is
particularly useful in this respect.

A brief, but almost certainly not complete, survey of known flare pumping models, from
the early stellar wind compression model by \citet{1982SvAL....8...86S} to the
very recent schemes listed in \citet{2019MNRAS.485.4676G}, suggests that none
of these authors themselves make detailed predictions of the expected IR light curve. The bipolar
outflow model by \citet{2012IAUS..287...93S} includes predicted periodic maser light
curves, but it is not clear how these relate to the pumping radiation. Interestingly,
these curves mostly show a slow rise and rapid decay of the maser, and so would not
obviously follow from the observed IR light curve in \citet{2018IAUS..336...37S}.

Flare-driving models have been compared \citep{2016A&A...588A..47V} with respect to their
ability to reproduce a periodic methanol maser light curve similar to that found in
G9.62+0.20E. To make this comparison, \citet{2016A&A...588A..47V} require light curves
for pumping or background radiation. They argue that the accretion-pulsation model
\citep{2013ApJ...769L..20I,2015ApJ...804L...2S} should have a light curve in the pumping
radiation that is qualitatively similar to a pulsating variable, for example a Cepheid,
since the pulsation is controlled by a similar opacity mechanism. Light curves for the
rotating spiral shock model by \citet{2014MNRAS.444..620P} were calculated by
\citet{2016A&A...588A..47V} for the cases of the stars alone, for the trailing
shock, and the combination of these. The trailing shock case looks qualitatively like
the maser light curves from \citet{2012IAUS..287...93S}: slow rise and rapid decay.

The model by \citet{2011AJ....141..152V} is different in that it produces the maser
flare via temporal change in the background radiation at the maser frequency, rather
than the IR pumping radiation. A model of similar type has been used to explain the
first known periodic H$_2$CO maser \citep{2010ApJ...717L.133A}. A binary protostellar
source is required with a colliding wind structure, and considerable eccentricity.
With the aid of additional formulae in \citet{2009MNRAS.398..961V}, there is enough
information to reconstruct the light-curve of the background radiation. Periodic 6.7-GHz
methanol masers appear to have several different forms for the light-curve, for example
\citet{2015MNRAS.448.2284S}, with the likely implication that the same is true of the
light curves of the pumping or background radiation.

\section{Modelling Considerations}
\label{s:model}

The present work approximates a fully time-dependent problem by a time-series
of time-independent solutions. There is therefore some timescale, for a given cloud, below which
the method becomes rather poor. The details relating to this timescale are discussed
first, in Section~\ref{ss:tscale}. In Sections~\ref{ss:ivd}-\ref{ss:bgv}, we discuss
additional important considerations related to the internal structure of the source
and the representation of the pumping models for external pumping radiation and background
intensity.

The code used in the present work is closely based on that used in \citetalias{2018MNRAS.477.2628G} and
\citetalias{2019MNRAS.486.4216G}. The main modification in the present work is a new capability to model
clouds that are non-uniform in the unsaturated inversion. This may be interpreted as either a cloud
of constant density with a variable pump rate, or constant pump rate within a
cloud of variable density. Clouds considered in this work either
have uniform physical conditions, or have an unsaturated inversion profile, $\delta \propto
1/r^2$, where $r$ is the radius drawn from the coordinate origin at the approximate
centre of the cloud. This dependence remains radial, even for clouds with a
non-zero deformation factor, $\Gamma$. As in \citetalias{2019MNRAS.486.4216G}, node distributions followed the shaping
equation,
\begin{equation}
(x^2 + y^2) e^\Gamma + z^2 e^{-2\Gamma} = 1,
\label{eq:shaping}
\end{equation}
where $(x,y,z)$ are Cartesian coordinates, such that $\Gamma<0$ ($\Gamma>0$) yields
an oblate (prolate) cloud.

\subsection{Applicable Timescales}
\label{ss:tscale}

A series of time-independent models is a good approximation to a time-dependent solution
provided that steady-state radiative transfer can be established on a time scale that is
much shorter than any timescale related to the maser pump. If the pumping
scheme is radiatively dominated, this corresponds to a timescale associated with 
the variation of an external radiation source
(either for the background or the pumping radiation). We establish a link between the flux of
pumping radiation and the optical depth multiplier of the cloud in Section~\ref{ss:pumpv}, so that
we may use a time-series of models of the same domain, but at different values of the
depth multiplier, provided that the evolution of that multiplier (or its
proxy, the pump rate) is suitably slow.

The most optimistic, or shortest, timescale on which steady state radiative transfer might be
established is the light crossing time of the cloud. For a cloud of radius
$r_{AU}$ astronomical units, this minimum time is $t_{lc} = 1000 r_{AU}$\,s. If the entire
pumping scheme relies only on transitions of modest optical depth, $\tau_p$, say up to order
$\sim 1$, it is a reasonable approximation to use $t_{lc}$, and the minimum representable
timescale of a time-series model will be no more than a few hours for maser clouds of reasonable
size.

We should, however, increase the minimum representable timescale to something considerably
larger than $t_{lc}$ on the following grounds. Firstly, the depth-multipliers of the maser
transition itself in the model take values up to 35: certainly not a modest depth. In the two-level
approximation, the pumping radiation is also linked to this single transition. However, in a multi-level
model of a real molecule, a complex pumping scheme might depend on transitions of even larger
optical depth - perhaps running up to values in the thousands. For example, the $5_{2,3}$ down to
$4_{1,4}$ transition of o-H$_2$O at 53.1\,$\umu$m has an optical depth of 5267 under conditions
corresponding the the largest 22-GHz inversion ($T_{kin}$=766\,K and $n_{H_2}$=6.7$\times$10$^{9}$\,cm$^{-3}$)
found in the parameter-space search modelling by \citet{2016MNRAS.456..374G}.
This transition plays an important role in the classic `collisional' pumping scheme for
the 22-GHz maser \citep{1973A&A....26..297D,mybook}. The geometry used in \citet{2016MNRAS.456..374G}
is different from the current work, and the clouds 
in that work have a scale size of 15\,AU. However,
optical depths of 350-1000 in the $5_{2,3}-4_{1,4}$ line might still be considered reasonable
for clouds considered typical in the present work.

In such a pumping scheme,
a diffusion solution is appropriate for a transition of very high optical depth, and then
the minimum representable time becomes the photon diffusion time of order $\tau_p t_{lc}$, that is
our original light crossing time multiplied by the optical depth in the pumping line. In this
case, minimum representable timescales of 12\,d apply for a 1\,AU radius cloud and 60\,d for a 5\,AU
radius cloud if a pumping scheme is controlled by a transition with an optical depth of
1000. Such timescales certainly impinge on the more rapid observed maser flares with rise
times of order weeks to months.

However, even if very optically thick transitions are important in a pumping scheme, the
situation is likely to be rather more optimistic than the estimates for a depth-1000 line
calculated above. The reason is the concept of effective scattering, quantified through the
scattering parameter. In the two-level case this is
\begin{equation}
\zeta = \frac{\Delta C_{u,l}}
             {\Delta C_{u,l} + g_u A_{u,l}},
\label{eq:scatparm}
\end{equation}
where $\Delta C_{u,l}$ is the nett downward collisional rate in the transition betweeen lower
level $l$ and upper level $u$, with statistical weight, $g_u$. The transition has a spontaneous
emission rate controlled by $A_{u,l}$, its Einstein A-coefficient. Effective scattering describes
a situation where a photon is formally absorbed in a transition, but there is a high probability
that a replacement photon will be spontaneously emitted into the same transition before the
energy of the original photon can be thermalized amongst the energy levels and thermal motions
of the molecules. Effective scattering then corresponds to the case where $A_{u,l}$ dominates
over the collisional rate and $\zeta$, from eq.(\ref{eq:scatparm}), is small.
Effective scattering reduces the optical depth from $\tau_p$ to the effective value of
$\sqrt{3\zeta} \tau_p$. This effective value is often significantly smaller than the value
based simply on the absorption coefficient. For example, for some optically thick 119-$\umu$m
transitions in OH, the $\sqrt{3\zeta}$ factor is 0.032 \citep{2007MNRAS.375..477G}. We
note that a transition that has high depth owing to a large A-value will, for the same reason,
likely have a small $\zeta$. Another useful example is the 45.1-$\umu$m pumping
transition in ortho-H$_2$O that supports inversion at 22\,GHz. The
45.1-$\umu$m transition has an Einstein A-value of 0.4205\,Hz, and an upper
level statistical weight of $g_u=$11. Under conditions close to those
for maximum maser depth (see Fig.~5 of \citet{2016MNRAS.456..374G}) the 
second-order downward
collisional rate coefficient 
is approximately 8.9$\times$10$^{-12}$\,cm$^{3}$\,s$^{-1}$. The corresponding
first-order collisional coefficient, at an H$_2$ number 
density of 6.7$\times$10$^9$\,cm$^{-3}$, is 0.06\,Hz. The net rate,
$\Delta C_{u,l}$ in eq.(\ref{eq:scatparm}) will be smaller. In fact, using
T$_{kin}$=750\,K, the upward rate coefficient is 0.0474\,Hz, yielding the
nett downward value of 0.012\,Hz. The value
of $\zeta$ for the 45.1-$\umu$m transition from eq.(\ref{eq:scatparm}) is
therefore 0.0026, yielding $\sqrt{3\zeta}$=0.088. This transition therefore also
has rather strong effective scattering, though not to the same extent as
the OH transition.

We suggest, given the examples above, that an order of magnitude reduction in the optically
thick timescales may be recovered on the basis of strong effective scattering in very thick
pumping lines, so that better estimates for the minimum representable timescales for the
model used in the present work are of order $1.2 r_{AU}$\,d.

In the case of variable background radiation, the limiting timescale is the light-crossing
time multiplied by the optical depth in the maser transition itself. The thickest models
considered have maser optical depths of 35 so, using the formula for the light-crossing
time above, the minimum representable timescale for variable background models is
$35000 r_{AU}$\,s, or 0.4\,d for a 1\,AU cloud. Somewhat shorter timescales are therefore
representable in variable background models.

\subsection{Clouds with Variable Unsaturated Inversion}
\label{ss:ivd}

The non-linear algebraic equations used to derive nodal solutions for the inversion
were derived in \citetalias{2018MNRAS.477.2628G}. The specific equation incorporating the finite-element
discretization used for computations is eq.(24) of that work. In the case of internal
variability, one cannot scale all nodal inversions by a single unsaturated value, so
eq.(3) of \citetalias{2018MNRAS.477.2628G} becomes incorrect. We write instead a version in terms of absolute
inversions, $\Delta(\vec{r})$, with unsaturated values, $\Delta_0(\vec{r})$, that
can now vary with position, $\vec{r}$. This more general form of \citetalias{2018MNRAS.477.2628G}, eq.(3),
is $\Delta(\vec{r}) = \Delta_0(\vec{r})/[1+\bar{j}(\vec{r})]$. We now divide this
expression by the maximum unsaturated inversion in the model, so that the left-hand side becomes
$\Delta(\vec{r}) /\Delta_{0,max} = \delta(\vec{r})$. A similar operation on the right-hand side
leaves
\begin{equation}
\delta(\vec{r}) = \delta_0(\vec{r})/[1+\bar{j}(\vec{r})].
\label{eq:modsat}
\end{equation}
Providing this new scaling is also applied to the gain coefficient in the radiation
transfer part of the problem, with a consequent change in the optical depth scale,
eq.(24) of \citetalias{2018MNRAS.477.2628G} takes the form,
\begin{equation}
\delta_i  - \delta_{0,i} \! \left[ \! 1 \!+\! \frac{i_{BG}}{4\pi} \!\!\sum_{q'=1}^{Q'} \!\frac{a_{q'}}{s_{q'}^2} \!\sum_{n=0}^\infty
                            \!\frac{1}{n!(n\!+\!1)^{1/2}} \! \left(
                                \sum_{j=1}^J \Phi_{j,q'} \delta_j \!
                                                        \right)^{\!\!\!n} \!
                   \right]^{-1} \!\!= 0,
\label{eq:globdisc}
\end{equation}
noting that the only formal modification is $\delta_{0,i}$, the scaled unsaturated
inversion at node $i$, multiplying the square-bracketed term. In eq.(\ref{eq:globdisc}), inversions at general
position $\vec{r}$ have been replaced by discretized versions specific to a node, indicated
by a subscript. We now apply a second, node-dependent, scaling to eq.(\ref{eq:globdisc}) by
dividing throughout by $\delta_0,i$. New fractional inversions $\delta'_{i} = \delta_i /\delta_{0,i}$
are now guaranteed to take values between 0 and 1, and are the variables solved for
computationally. The slightly modified final form of eq.(\ref{eq:globdisc}) is
\begin{equation}
\delta'_i  -  \! \left[ \! 1 \!+\! \frac{i_{BG}}{4\pi} \!\!\sum_{q'=1}^{Q'} \!\frac{a_{q'}}{s_{q'}^2} \!\sum_{n=0}^\infty
                            \!\frac{1}{n!(n\!+\!1)^{1/2}} \! \left(
                                \sum_{j=1}^J \delta_{0,j} \Phi_{j,q'} \delta'_j \!
                                                        \right)^{\!\!\!n} \!
               \right]^{-1} \!\!= 0.
\label{eq:satf}
\end{equation}
Equation~\ref{eq:satf} shows that the same set of pre-computed saturation coefficients
used in Papers~1 and 2, the
$\Phi_{j,q'}$, can be used with variable unsaturated inversion too, provided that
 each coefficient is multiplied by
the appropriate scaled unsaturated inversion.

A further computational note is that eq.(\ref{eq:satf}) has an optical depth scale based
on the maximum nodal unsaturated inversion in the model. This scaling is rather thick compared to the
typical behaviour of a uniform cloud. It is computationally convenient to have global
optical depth multipliers that give approximately similar levels of saturation for an
internally variable cloud and one with uniform unsaturated inversions. A multiplicative bias is therefore applied
to the depth scale. In the present work, this was the mean of the scaled nodal unsaturated inversions. However,
this might not be appropriate for all internal variability functions, and we
prefer scaling by the maximum unsaturated inversion, and the adoption of a separate bias, to a default
scaling by the mean unsaturated inversion.

\subsection{Notes on Pumping Variability}
\label{ss:pumpv}

In a two-level model of the type used in the present work, varying the pump rate of
the maser (or to be more precise, the ratio of the pump rate to the loss rate) is
equivalent to varying the optical depth of the model. This assertion is justified
as follows: Variation in the pump rate changes the unsaturated inversion. As we have
seen in Section~\ref{ss:ivd} above, the unsaturated inversion controls the optical
depth scale. In a uniform model, this will be via a global multiplier; in an internally variable
model, there will be a global multiplier that is in turn multiplied by a node-dependent
unsaturated inversion. Overall, however, a time variable pump rate corresponds to 
a time-variable depth multiplier.

Computationally, the conclusion from the paragraph above is vital: a
sequence of solutions, for a particular computational domain, at a number of different
values of the depth multiplier can be used to represent a sequence of snapshots in time with
different values of the pump rate. There are some provisos to this conclusion.
The most important proviso is that the time-dependence
of the pump rate is slow compared to any processes, including radiative transfer,
within the cloud (or domain). This point is discussed in more detail in terms
of the minimum representable timescale in Section~\ref{ss:tscale}. A second proviso
is that a suitable function can be found to represent the time dependence,
preferably with observational constraint. Once a suitable function has been
selected, inversions at all nodes of the domain can be generated as a smooth
function of time (as a proxy for depth multiplier) via interpolation.

\subsection{Internal Variability of the Pump}
\label{ss:pumpspace}

Whilst it is true that the inversion at any node is proportional to its local pump rate, and the
pump rate is proportional to the optical depth multiplier of the model, as justified
in Section~\ref{ss:pumpv}, it is not trivial to justify
extending this proportionality all the way to the energy density of infra-red radiation
provided by a protostar or other source. For some key pumping transition, the pump
rate will be proportional to the mean intensity in that transition at the node. However,
that local mean intensity is not trivially linked to what might be incident on the surface
of the cloud. 

We have introduced, in Section~\ref{ss:tscale}, the idea of an optically thick pumping line in the
effective scattering approximation. Solution of the radiation diffusion equation in this
limit yields an expression for the mean intensity at a node where the pumping line depth is
$\tau_p$:
\begin{equation}
\bar{J}(\tau_p) = \bar{J}(0) e^{-\sqrt{3\zeta} \tau_p} + B_{\nu}(T_K),
\label{eq:diffnsol}
\end{equation}
where $\bar{J}(0)$ is the surface mean intensity that follows the driving light curve, and
$T_K$ is the kinetic temperature local to the node. 

We suggest that the two cloud types used
in this work should have the following interpretations: The clouds with uniform unsaturated
inversions should be interpreted as clouds with a realistic density profile, that rises towards
the cloud centre, but is compensated by a falling energy density of pumping radiation,
predicted by eq.(~\ref{eq:diffnsol}). Obviously this is only an approximation, since
the functional forms for the physical density and the energy density of pumping radiation are
most unlikely to compensate each other exactly. In this interpretation,
The internally uniform clouds represent a thicker regime in respect of
the pumping radiation, where $\sqrt{3\zeta} \tau_p$ can
rise to perhaps 5-10, and $\tau_p$ can be much larger. The second type of cloud, with
variable unsaturated inversion, should be interpreted again as a cloud with a centrally
condensed density profile, but where we now assume that
the first term in eq.(~\ref{eq:diffnsol}) dominates the pumping. This implies that the effective
optical depth in the pumping line must be small, even if $\tau_p$ is substantial. 
The internally variable clouds therefore correspond to the case of optically thinner
pumping lines, where $\sqrt{3\zeta} \tau_p$ is small.

\subsection{Notes on Background Variability}
\label{ss:bgv}

The specific intensity of the background radiation, $i_{BG}$, amplified by the maser is a
parameter that appears in eq.(\ref{eq:satf}). Therefore, provided this parameter varies
slowly compared to the minimum representable timescale for the cloud, we may adopt
a time-series procedure similar to that used for variation in the pumping radiation,
but we now use a series of models with different background intensities to study
the light curve, rather than models with different depth multipliers. The depth
multiplier is generally kept constant in models where the background varies.
In the present work, the background specific intensity is considered to be uniform
over solid angle. This flare mechanism has been intoduced by
\citet{2011AJ....141..152V,2016A&A...588A..47V}. However, it is an unlikely explanation
for the periodic G107.298+5.63 system, since a variable background would be expected
to produce correlated H$_2$O 22\,GHz and methanol 6.7\,GHz emission, rather than
the observed anti-correlation that 
strongly favours a variable pumping explanation \citep{2016MNRAS.459L..56S}.

\section{Domains}
\label{s:domains}

In order to cover a reasonable range of models, we considered three standard domains
with deformation factors of $\Gamma = -0.6,0.0,+0.6$ corresponding, repectively, to 
models that are substantially oblate, quasi-spherical and substantially prolate. 
As in \citetalias{2019MNRAS.486.4216G}, deformations were applied to originally spherical point distributions
via eq.(\ref{eq:shaping}). Views
of the two domains with $\Gamma \neq 0$ appear in Fig.~1 of \citetalias{2019MNRAS.486.4216G}, and a 
somewhat smaller quasi-spherical domain appears in Fig.~1 of \citetalias{2018MNRAS.477.2628G}. We do not
repeat plots of the domain structure here.

All three domains may be presented in the form with internally uniform conditions, and in a version
where unsaturated inversions, $\delta_{0,i}$, at the nodal points follow the formula
\begin{equation}
\delta_{0,i} = \delta_{out} (r_i/r_{out})^{-2} ,
\label{eq:densfunc}
\end{equation}
where $\delta_{out} = 1.0$ is the unsaturated inversion at the node with the largest radius, $r_{out}$,
measured from the origin of the domain.
The variable unsaturated inversion models were designed to produce a core-halo structure that
is common in many observed masers at VLBI resolution.
The extent to which this was successful or not may be judged from the results
in Section~\ref{s:results}.
The number of nodes per domain was similar to those used in \citetalias{2019MNRAS.486.4216G}, with approximately
250 of an original list of 300 nodes surviving the triangulation process.

The same ray-tracing algorithm was used here as in \citetalias{2018MNRAS.477.2628G} 
and \citetalias{2019MNRAS.486.4216G}, so that solutions
for the inversions involved tracing 
1442 rays from points on a celestial sphere-style source to each target node
of the domain.
The celestial sphere source had a standard specific intensity of $i_{BG}=1.0\times 10^{-6}$, 
in cases where the maser pump was varied, and a range of $10^{-9}$ to $10^{-5}$ in the
case of a variable background, where $i_{BG}=I_{BG}/I_{sat}$, and $I_{sat}$
is the saturation intensity of the maser. The upper limit of the variable background corresponds
to the background intensity used 
in \citetalias{2018MNRAS.477.2628G} and \citetalias{2019MNRAS.486.4216G}. The background and pumping radiation is taken to be isotropic in
all cases.

\section{Results}
\label{s:results}

First, in Section~\ref{ss:ilc}, we introduce the functional forms that we will use as models
for the light curves of pumping and background radiation, before briefly demonstrating
the main effects of changing the model from one with uniform physical conditions, to
one with variable unsaturated inversions (Section~\ref{ss:nodalsol}).

The main results, in the form of formal radiative transfer solutions, seen as images
and spectra by an observer,
appear in Section~\ref{ss:resvarpump} and Section~\ref{ss:resvarback}, where
we discuss, respectively, the effects of pumping and background radiation on the maser, in
terms of saturation, cloud shape and the observer's viewpoint. This work is extended to a
consideration of light curves with statistical information in Section~\ref{s:reslightcurve}.

\subsection{Input Light Curves}
\label{ss:ilc}

In the rest of this work, we use the following driving light curves to represent
the time variation of either the pumping or background radiation: function $D_0$ is
a default sinusoid, $1+\sin(x)$; function $D_1$ is a digitized version of the observed IR light
curve from the work of \citet{2018IAUS..336...37S}; function $D_2$ is a theoretical Cepheid
variable light curve from the top left panel of Figure~1 from \citet{2017MNRAS.466.2805B}, and is used 
to represent the accretion-pulsation
model, following the suggestion of \citet{2016A&A...588A..47V}; functions $D_3$-$D_5$
are respective digitized forms of the light curves used by \citet{2016A&A...588A..47V} to represent pumping
by the stars alone, by the trailing shock and by the sum of these sources following
the spiral shock model by \citet{2014MNRAS.444..620P}. In a little more detail, this model
consists of a primary and secondary star in a circular binary orbit within a circumbinary
disc. The primary has a trailing shock that heats gas and dust (see Fig.~2 of \citealt{2016A&A...588A..47V}).
The light curve follows the illumination of a point on the inner edge of the circumbinary
disc by the stars, the shocked material, or both of these sources. Finally, function $D_6$ is the
light curve for background radiation predicted by the colliding-wind eccentric binary
model. The decay is constructed from equations in \citet{2009MNRAS.398..961V}, but the rise has
been modelled by a simple exponential, as \citet{2009MNRAS.398..961V} do not provide a
functional form for this part of the light curve. The recently published light curve from \citet{2019PASJ..tmp..129U} is
discussed briefly in Section~\ref{discuss}.

\begin{figure}
  \includegraphics[bb=320 100 560 570, scale=0.45,angle=0]{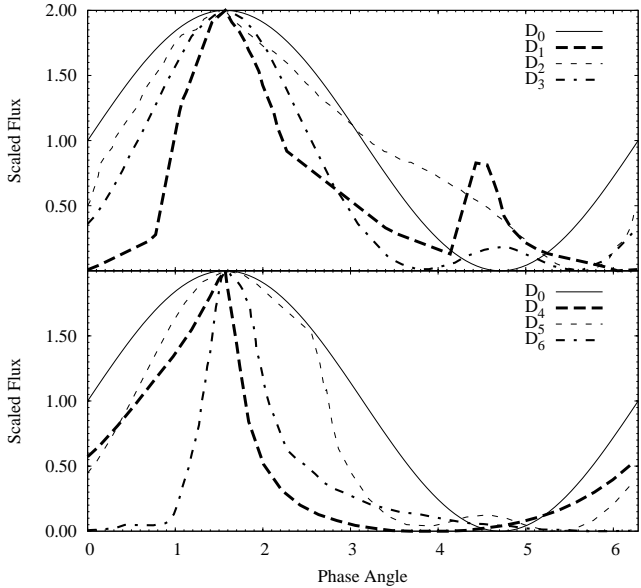}
  \caption{Driving functions used in the flare modelling in later sections. The
top panel shows functions $D_1 - D_3$ (see text) compared to the default
sinusoid, $1+\sin(x)$, as $D_0$. The lower panel compares $D_0$ with driving
functions $D_4 - D_6$. All functions are scaled to an amplitude of 1 with the maximum 
occurring at $\pi/2$\,radians.}
\label{f:drivers}
\end{figure}

All the functions, $D_0$-$D_6$, are drawn in Fig.~\ref{f:drivers} with equal $y$-axis range and with
maximum light at a phase of $\pi /2$. Where functions have been digitized, this
process was carried out with the WebPlotDigitizer tool, by 
Ankit Rohatgi\footnote{https://automeris.io/WebPlotDigitizer}. Where
necessary, continuous functions were
reconstructed from the digitized data via a cubic splining process, based on the
functions {\sc spline} and {\sc splint} \citep{1992nrfa.book.....P}. The original Cepheid
light curve was presented with a $y$-axis in magnitudes, and this curve has been re-plotted
here on a linear scale.
In this work, we do not specify a particular functional form for the IR light-curve of a source
of the accretion-burst type, for example S255 NIRS3, on the grounds that present data are too
sparse to generate anything that might be regarded as typical.

\subsection{Nodal Solutions}
\label{ss:nodalsol}

Nodal solutions were generated in a broadly similar manner to those 
in \citetalias{2018MNRAS.477.2628G} and \citetalias{2019MNRAS.486.4216G}. The
interesting point in the present work is that the sequence of solutions at increasing optical
depth now represent different levels of pumping radiation, and therefore potentially correspond
to different times in a radiation-driven pumping cycle. To limit the overall complexity of
the parameter space, we restrict the domains used in the present work to one oblate, one
prolate and one spherical domain. These have the respective deformation factors
-0.6, 0.6 and 0.0 from eq.(\ref{eq:shaping}).
To allow us also to model the effects
of a variable background, several sets of solutions were generated with background levels, in
saturation units, ranging from 10$^{-9}$-10$^{-5}$ for the prolate and oblate clouds.

All solutions in the present work were computed using the non-linear Orthomin(K) algorithm
\citep{2001ApMCo.124..351C}, with the order $K=2$. For each domain considered, solutions were computed
for optical depth multipliers of 0.1 to at least 30.0 in steps of 0.1. All solutions assumed
complete velocity redistribution, and we note that the levels of saturation reached for
the highest depth multipliers ($\tau \sim 30.0$), correspoding to actual maser optical depths
of order 60 on some lines of sight, are almost certainly at the upper limits of the
validity of this approximation, even for rapid redistribution mediated by the pumping
radiation \citep{1988rmgm.conf..339F,1994A&A...282..213F,mybook}.

\begin{figure*}
  \includegraphics[bb=0 80 595 780, scale=0.80,angle=0]{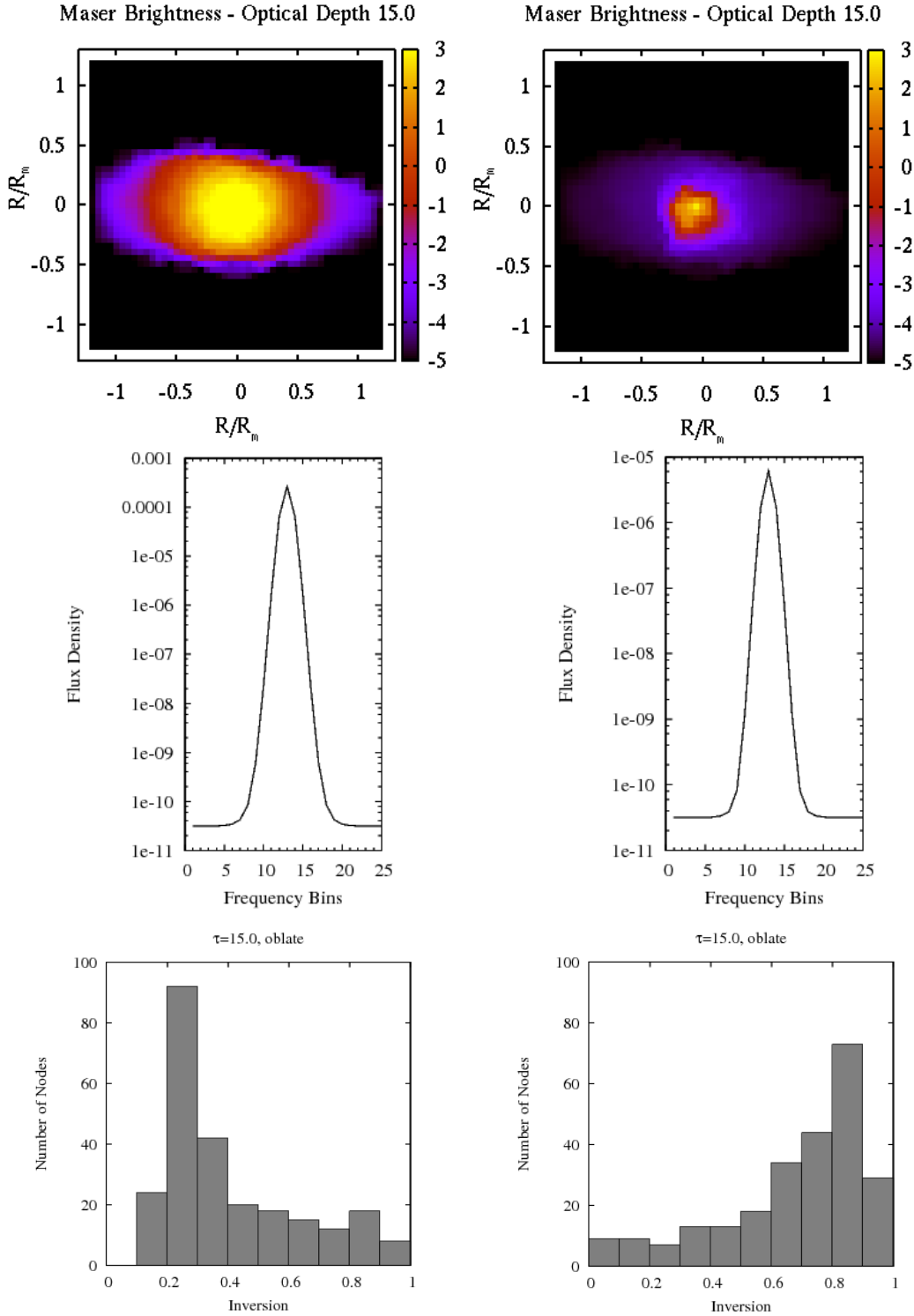}
  \caption{Images, spectra and saturation histograms for $\Gamma=-0.6$
oblate clouds with uniform conditions (left-hand
column) and a $1/r^2$ profile of the
unsaturated inversion (right-hand column). Both datasets use the same
geometric distribution of nodes. The spectra and maps are as viewed by an observer at
polar angle = azimuthal angle = $\pi/2$.}
\label{f:deepclouds}
\end{figure*}

To demonstrate the difference between the uniform and internally variable clouds, we show
in Fig.~\ref{f:deepclouds} images and spectra of both cloud styles for the oblate domain, viewed
edge on, together with histograms that show the level of saturation in the model through
the occupancy of bins that represent remaining fractional inversion. Both examples have
a moderate optical depth multiplier of $\tau = 15.0$, and were amplified from a background
of $i_{BG}=10^{-5}$.

The internally variable clouds are perhaps a reasonable representation of a core-halo structure
that is sometimes present in 6.7-GHz methanol maser sources. From Fig.~\ref{f:deepclouds},
the outstanding features of adopting a cloud with 
variable unsaturated inversion, weighted such that inversions are highest
towards the centre, is to reduce the angular size of the brightest region of the image. The
spectrum is narrower, but has a lower flux density, than the uniform case. The saturation
profile is significantly different too: the uniform cloud is arguably more highly saturated
generally, with more nodes in the $0.1-0.2$ inversion bin than in any other. However, the internally
variable cloud has a significant population in the most saturated bin $0.0-0.1$ of
the original inversion remaining, whilst this bin is empty in the uniform case.

\subsection{Variable Pumping Radiation}
\label{ss:resvarpump}

We begin by studying the effect of saturation by applying functions $D_0 - D_5$ to
represent the time-series in optical depth multiplier, as a proxy for the maser 
pump rate, as discussed in Section~\ref{ss:pumpv}, with
the caveats raised in Section~\ref{ss:pumpspace}. These functions were applied in the
form,
\begin{equation}
\tau (t) = \tau_{min} + (1/2) \Delta \tau D_n (t),
\label{eq:pumpthing}
\end{equation}
where $\Delta \tau = \tau_{max}-\tau_{min}$, and $\tau_{min}$ ($\tau_{max}$) is the optical depth multiplier
of the thinnest (thickest) nodal solution employed, and $n$ is in the range 0-5. Apart from
the function index, $n$, there are two parameters in eq.(\ref{eq:pumpthing}): the minimum
optical depth multiplier, $\tau_{min}$, and the multiplier range, $\Delta \tau$. Suitable values
of these parameters should be informed by observation and available data. The multiplier
range is perhaps the easier parameter to constrain via the original y-axis ranges of functions
$D_1 - D_5$. Excluding function $D_2$, where the original data is in magnitudes, leading to a
flux ratio (of 1.629), the remaining functions have changes in the flux of pumping radiation
ranging from 0.94 ($D_4$) to 3.22 ($D_1$). As these changes are proportional to changes in
the optical depth scale of the domain, it is reasonable to set these, or the rounded figures
of 0.9-3.5 as the bounds of $\Delta \tau$. Given that the longest maser gain lengths in a scaled
cloud are closer to 2 than 1 (diameter rather than radius) then
if the amplification is entirely unsaturated, these
figures would lead to maser intensity ratios between maximum and minimum light, along a given
line of sight, of 6.05 and 1100. These figures are perhaps a
poor guide anyway, as there are no light curves available at the wavelength range that is
known to be important for pumping. We typically used $\Delta \tau =5.0$, and occasionally
larger figures on the grounds that observed maser flares are known with variability indices
considerably in excess of 1000.

For each model, two formal solutions were computed for every nodal solution, or
optical depth multiplier, available. The first of these places the observer close to
the optimum viewing position for the prolate and oblate clouds. The second uses a
randomly chosen viewpoint. The reason for choosing the first type is that observations
are obviously biased towards the detection of the brightest masers, so we are
quite likely to be observing objects close to optimally oriented to our line of
sight. The second type of formal solution gives us an idea of the sort of flare that
might result from an average cloud, selected from an ensemble of similar objects.

Light curves were constructed by taking one of the observable parameters from the
formal solutions and computing its response to the driving in depth from
eq.(\ref{eq:pumpthing}). In detail, this was carried out by fitting a cubic natural
spline through the parameter values at the known depths.

\subsection{Variable Background Radiation}
\label{ss:resvarback}

The operations for variation of the background radiation follow a broadly similar approach
to those used in Section~\ref{ss:resvarpump}. However, there are important differences. Firstly,
we apply only the default sinusoid and function $D_6$ as driving functions, since only $D_6$
is based on a model where the background, rather than the pump, is varied. The process is easier
in the sense that the parameter that is varying in time is simply a parameter in eq.(\ref{eq:globdisc}),
and no complicated justification is required for its use. A significant complication in the variable
background case, however, is that the driver in the background analogue of eq.(\ref{eq:pumpthing})
\begin{equation}
i_{BG} (t) = i_{BG,min} + (1/2) \Delta i_{BG} D_6 (t),
\label{eq:bgthing}
\end{equation}
now demands a response from a number of different models, each computed with a different value of
$i_{BG}$. Therefore, modelling flares from background variation is computationally considerably
more expensive.

\section{Variable Pumping Results}
\label{s:reslightcurve}

\begin{figure}
  \includegraphics[bb=320 100 560 570, scale=0.45,angle=0]{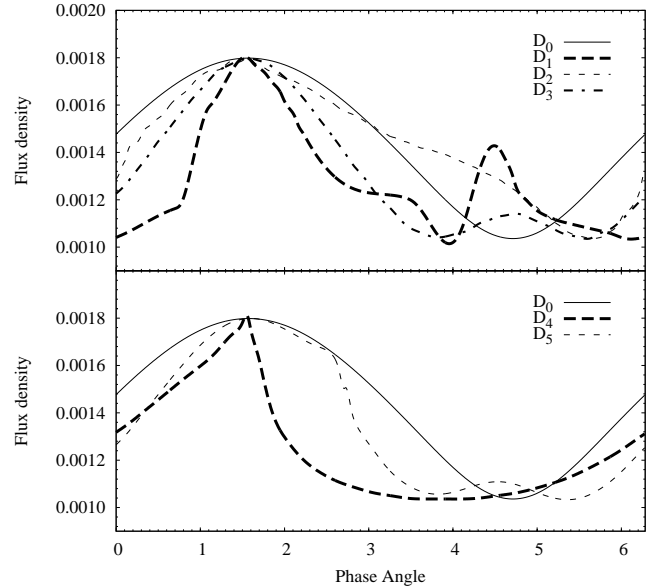}
  \caption{Maser light curves resulting from driving functions $D_0$ to $D_3$
(upper panel and $D_4$, $D_5$ with $D_0$ (lower panel) applied
to a $\Gamma=0.6$ prolate cloud. Optical depth multipliers
in this model ranged from 10.0 to 15.0. The observer is in the optimum viewing
position.}
\label{f:fig3}
\end{figure}

We plot, in Fig.~\ref{f:fig3}, the flare light curves, over one period, that result from driving
with pump functions $D_0-D_5$ for a prolate cloud with $\Gamma=0.6$, with the observer placed in the optimum
position (viewing the cloud along the $z$-axis close to the long axis of the cloud). The
cloud style is uniform in unsaturated inversion.
The parameters of eq.(\ref{eq:pumpthing}) in this case are
$\tau_{min}=10.0$ and $\Delta \tau=5.0$, corresponding to a regime that is already characteristic
of quite strong saturation. Throughout Section~\ref{s:reslightcurve}, the $y$ axes of graphs are
shown in flux density, based on a background specific intensity that is scaled to the saturation
intensity of the maser transition. Flux density is likely the
best parameter to plot in view of the prominent role of
single-dish monitoring in repetitive flare studies. An explanation of the flux-density scale
appears in Appendix~\ref{a:fdunits}. Owing to the nature of the saturation intensity, and
particularly the fact that this quantity may itself depend on the energy density of pumping 
radiation, a conversion formula to units of Jy at a typical source distance, and for
a particular maser transition, for example, is
not trivial. Such a conversion is also discussed in detail
in Appendix~\ref{a:fdunits}.
The variability index of the maser light curve, defined as in the Introduction as $F_{pk}/F_{qui}$
is 1.74 in Fig.~\ref{f:fig3}. The same parameter for the driving functions, replacing
flux densities with values of the depth multiplier, is 1.5. Although the variability index
of the maser is significantly larger here, the light curve in Fig.~\ref{f:fig3} demonstrates the
smoothing effect of saturation: as exponential growth gives way to a functional
form that is closer to linear at high flux density, the maser responds more weakly to the
driving function, smoothing particularly the high flux density parts of the light curves.
Note that, compared to Fig.~\ref{f:drivers}, secondary
features at low flux density, such as the small peak in function $D_1$, are exaggerated. 

\begin{figure}
  \includegraphics[bb=320 100 560 570, scale=0.45,angle=0]{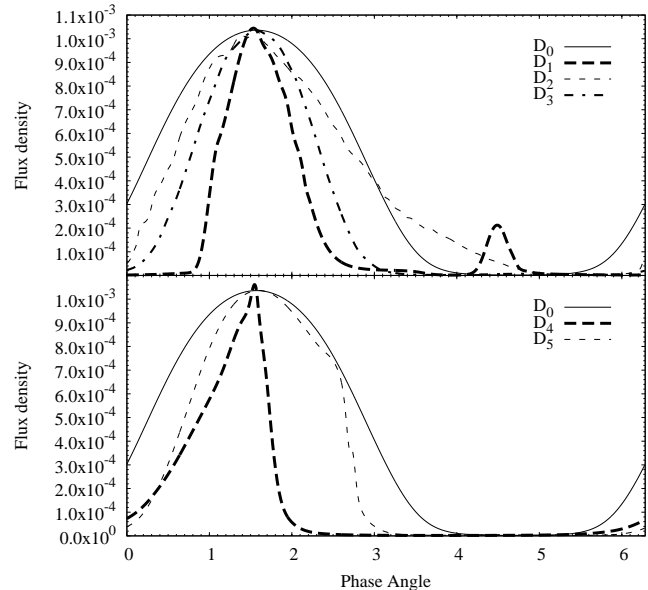}
  \caption{As for Fig.~\ref{f:fig3} but with an optical depth
multiplier range of 5.0 to 10.0.}
\label{f:fig4}
\end{figure}

Fig.~\ref{f:fig4} depicts a scenario where growth is
less saturated. Again the amplitude in depth parameter is 5.0, but now the
optical depth range lower and upper
limits are 5.0 and 10.0, respectively, so that the upper limit flux density here corresponds
to the lower limit in Fig.~\ref{f:fig3}.
The variability index of the flare in Fig.~\ref{f:fig4} is 502, compared with $1.74$
in Fig.~\ref{f:fig3}. The weaker parts of the light curve are now compressed down to the
$x$-axis, and the duty cycle, defined as the
fraction of the period over which the light curve is above the
mean of its maximum and minimum values, is reduced compared to the driving functions.

\begin{figure}
  \includegraphics[bb=320 100 560 570, scale=0.45,angle=0]{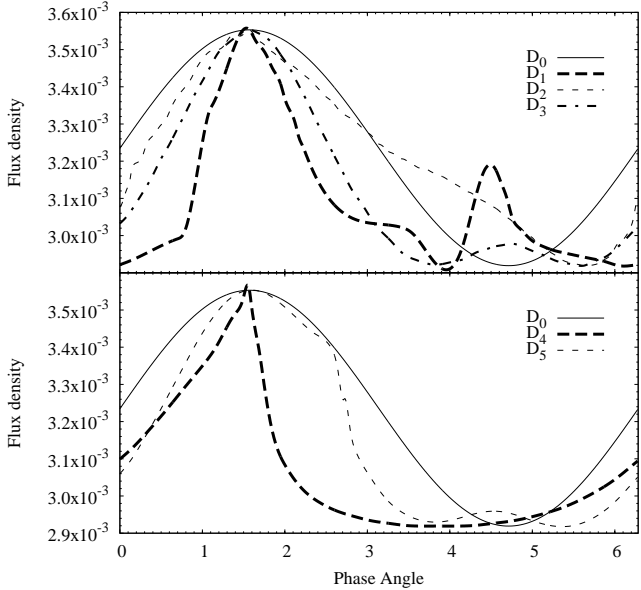}
  \caption{As for Fig.~\ref{f:fig3} but with an optical depth
multiplier range of 24.0 to 29.0.}
\label{f:fig5}
\end{figure}

As a last example of the effects of saturation, we show in Fig.~\ref{f:fig5} a situation with
the same amplitude in depth (or mean intensity of pumping radiation) of 5.0, but with a 
minimum depth multiplier of 24.0, corresponding to extreme saturation in the context of
this model.
The compression of the response is more extreme than in Fig.~\ref{f:fig3}, with a variability
index of the `flare' now only 1.22 (compared to
1.208 for the driving functions). In terms of shape, the curves in Fig.~\ref{f:fig5} more strongly
resemble the driving functions plotted in Fig.~\ref{f:drivers} than the case of moderate
saturation in Fig.~\ref{f:fig3}, where the high flux density parts of the response are broadened, or
of the weak saturation case of Fig.~\ref{f:fig4}, where the high flux density parts of the response
are narrowed in a high variability index flare. This behaviour is in accord with the linear gain
expected of highly saturated masers.

\subsection{Effect of Viewpoint}
\label{ss:viewpoint}

\begin{figure}
  \includegraphics[bb=320 100 560 570, scale=0.45,angle=0]{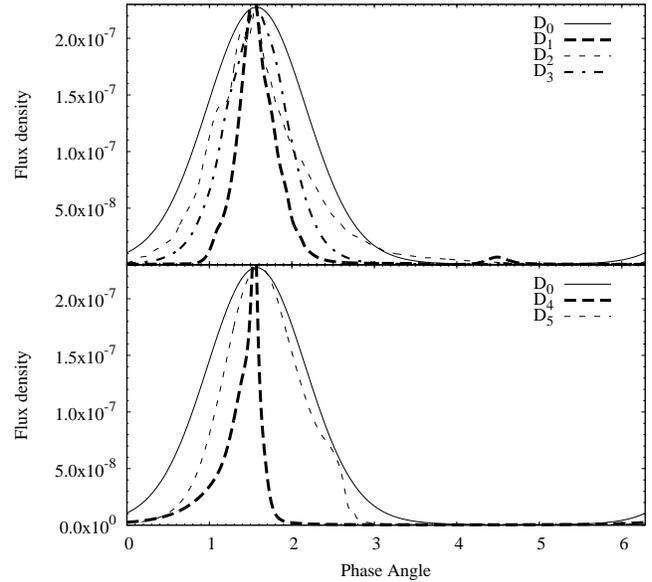}
  \caption{As for Fig.~\ref{f:fig4} but with the observer's viewpoint
shifted to $(\theta,\phi)=(1.081,3.465)$\,radians.}
\label{f:typical}
\end{figure}

In this section, the observer's viewpoint is moved to a `typical' position instead of the view close
to the long-axis of the domain. In spherical polar coordinates, the chosen position has polar angle of
1.081 radian, and an azimuthal angle of 3.465 radian. Light curves that result from a driving range
in depth from 5.0-10.0, analagous to Fig.~\ref{f:fig4}, are plotted in Fig.~\ref{f:typical}.
We defer a discussion of the statistics related to a large number of random viewpoints
until Section~\ref{s:stats}. The effect of moving away from the optimum viewpoint is
highly significant. The variability index in Fig.~\ref{f:typical} is 580, as opposed to
502 in Fig.~\ref{f:fig4}, indicating reduced saturation. However, the shapes of the light
curves indicate this more visually: there is no broadening of the curves in the higher flux
density regime, indicating little saturation even at the highest flux densities attained
in Fig.~\ref{f:typical}. A glance at the $y$ axis shows that the highest flux density
attained is only 2.28$\times$10$^{-7}$, a factor of 4600 below that obtained in the optimum
orientation, and in fact almost an order of magnitude below the minimum achieved
in Fig.~\ref{f:fig4}.

\subsection{Extreme Flares}
\label{ss:flares}

\begin{figure}
  \includegraphics[bb=320 100 560 570, scale=0.45,angle=0]{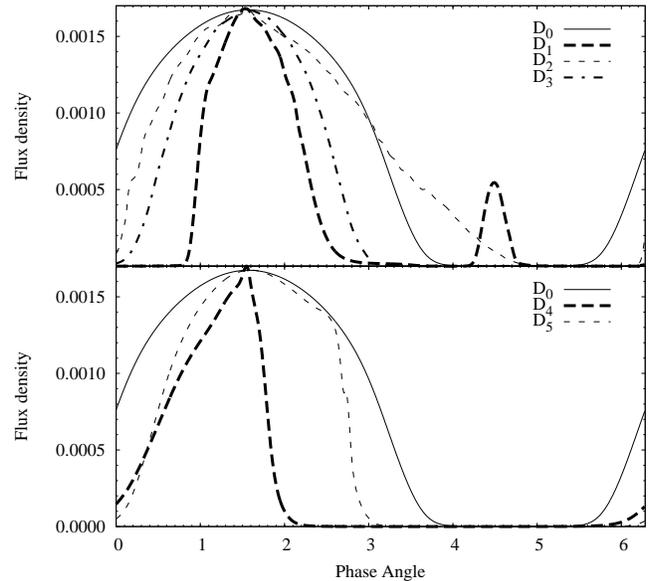}
  \caption{As for Fig.~\ref{f:fig3} but with an optical depth
multiplier range of 4.0 to 14.0.}
\label{f:extreme}
\end{figure}

From the work in Section~\ref{s:reslightcurve} so far, it is clear that the way to get a high
amplitude flare from variable IR
pumping is to introduce a large $\Delta \tau$ in eq.(\ref{eq:pumpthing}), and to begin
from a rather low value of $\tau_{min}$, so that the range includes a substantial amount of
unsaturated growth. However, another consideration is observability: a very large amplitude in
the maser response is no good if the flare maximum is still too weak to see. A good compromise,
perhaps, is to set $\tau_{max}$ in the moderately saturated regime, whilst $\tau_{min}$
is in the range that is essentially unsaturated. In Fig.~\ref{f:extreme} we show the
light-curves resulting from the optimally oriented prolate cloud with $\tau_{min}=4.0$ and
$\Delta \tau = 10.0$.
This set of curves maintains a maximum flux density that is only 0.47 times the maximum
at the highly saturated depth multiplier of 29.0, whilst starting from a depth multiplier
low enough to yield a variability index of 1.65$\times$10$^{4}$. Even the $D_0$ function in
Fig.~\ref{f:extreme} has a broad range that appears indistinguishable from the baseline
on this linear graph. However, saturation near maser maximum prevents the peak narrowing
to a form with a much smaller duty cycle.

\subsection{Effect of Cloud Shape}
\label{ss:cloudshape}

\begin{figure}
  \includegraphics[bb=60 50 400 440,scale=0.46,angle=0]{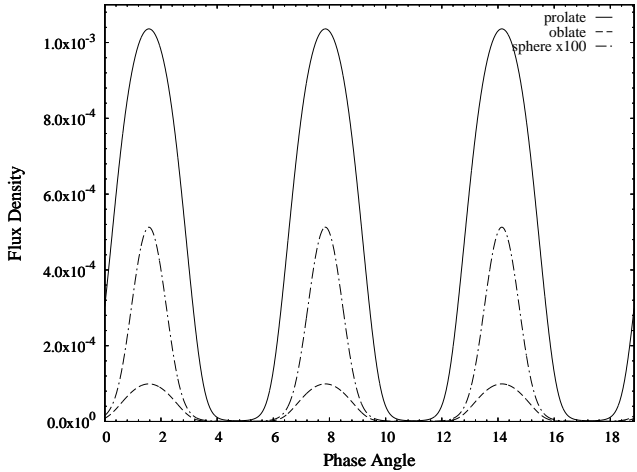}
  \caption{Maser light curves for the prolate (solid line), oblate (dashed line) and 
pseudo-spherical (chained line) clouds driven
by function $D_0$ over three cycles with maser depth range 5-10. Note that the plotted
light curve of the spherical cloud is multiplied by 100 to make it distinct from the
$x$ axis.}
\label{f:tricycle}
\end{figure}

We compare here results for a pseudo-spherical cloud, and 
an oblate cloud with the same magnitude of distortion as the
prolate cloud studied in Section~\ref{ss:viewpoint} and most of the more general 
Section~\ref{s:reslightcurve}. To reduce complexity, only the default driving
function $D_0$ is considered. We plot the maser response to this function for
the three cloud shapes, over three periods, in Fig.~\ref{f:tricycle}. The range for
the maser depth multiplier $\tau$ is 5-10, as in Fig.~\ref{f:fig4}.
The observer's position is `optimum' for the prolate and oblate clouds.
For an oblate cloud, this must lie close to the $xy$ plane, and we have used
$(\theta,\phi)=(\pi/2,\pi/2)$. We have used the same
`typical' position as in Section~\ref{ss:viewpoint} for the pseudo-spherical object. 
We note from \citetalias{2018MNRAS.477.2628G} that variation
of the viewing point for clouds of this type may alter the maser output by a 
factor of up to $\sim$3, but the maximum flux density of the spherical cloud
in Fig.~\ref{f:tricycle} is more than two orders of magnitude lower than that of the
optimally viewed prolate cloud of the same volume. The respective variability indices
for the prolate, oblate and pseudo-spherical clouds in Fig.~\ref{f:tricycle} are
502, 2930 and 1760. The variability index of the prolate cloud is the smallest of
the three because it has the highest minimum in Fig.~\ref{f:tricycle}, due to its
geometry, whilst its maximum is significantly affected by saturation. The minima of all three curves
merge with the $x$ axis in Fig.~\ref{f:tricycle} due to the linear flux density scale.

Over the partially saturated
depth multiplier range of 10.0-15.0, as used in Fig.~\ref{f:fig3}, the maximum flux density
achieved with the oblate cloud was 3.0$\times$10$^{-4}$, a factor of 5.99 smaller than in
the prolate case. The pseudo-spherical object was weaker still, with a maximum flux
density of 7.05$\times$10$^{-5}$. Light curves for this depth multiplier range are
not plotted, but their
respective variability indices
were 1.74, 3.03 and 13.52 for the prolate, oblate and pseudo-spherical clouds,
implying that the light curves of the oblate and spherical clouds are somewhat less 
affected by saturation. The shape of the light curve for the 
oblate cloud is not significantly different from the prolate case, with the strong saturation flattening towards
higher optical depth multipliers, as in Fig.~\ref{f:fig3}.

Even at the ultimate saturation level modelled, $\tau=30$, the optimally viewed prolate cloud, with
a flux density of 3.67$\times$10$^{-3}$ is still 4.83 times brighter than the optimally viewed oblate
cloud, and 12.2 times brighter than the spherical cloud.

When shifted to the typical position, and here the same position is used for all types
of cloud, we find that the maser flux density follows a  qualitatively similar family of curves
to those in Fig.~\ref{f:typical}. However, the maximum flux density order is now reversed with
the spherical cloud at 5.22$\times$10$^{-6}$ being 19.84 times brighter than the oblate cloud
and 22.93 times brighter than the prolate cloud. In terms of variability index the order
is spherical, prolate, oblate with respective indices 1760, 502 (as in Fig.~\ref{f:typical})
and 341, indicating that the pseudo-spherical cloud is also the least affected by saturation
from this viewpoint. However, the oblate cloud has the largest improvement in variability index
by dropping the depth range to 5-10 from 10-15 from the optimum viewpoint.

\subsection{Clouds with Internal Structure}
\label{ss:tvdcloud}

\begin{figure}
  \includegraphics[bb=320 100 560 570, scale=0.45,angle=0]{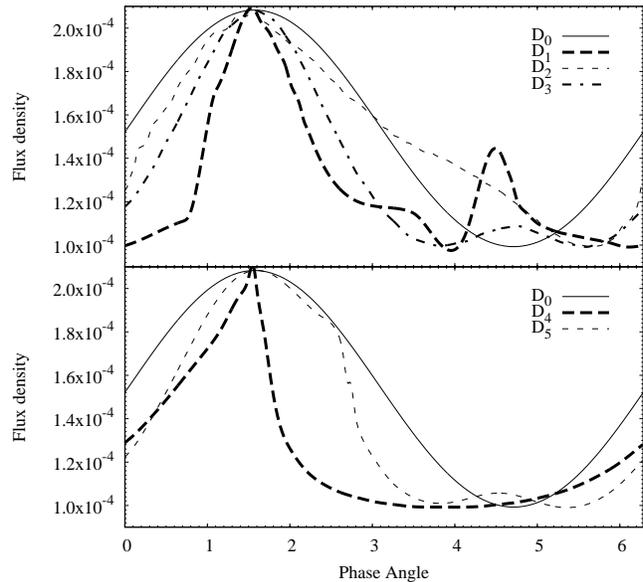}
  \caption{As Fig~\ref{f:fig3} but for a cloud with an unsaturated inversion profile
that follows a $1/r^2$ function, where $r$ is the radial distance from the cloud centre.}
\label{f:fig_tvd}
\end{figure}

The clouds introduced here have an unsaturated inversion that increases towards the centre of the 
cloud. As discussed in Section~\ref{ss:pumpspace}, this situation most likely corresponds to a case where
the optical depth to the pumping radiation is fairly low, and a real increase in the number
density of the maser molecule more than outweighs any decrease in the energy density of the
pumping radiation towards the cloud centre. In Fig.~\ref{f:fig_tvd}, we show a situation 
comparable to that in Fig.~\ref{f:fig3} in terms of the minimum and maximum optical depth
multipliers considered, and the optimally placed observer.

The effects of moving to a variable density cloud are not particularly profound,
at least in terms of flux density. The curves 
in Fig.~\ref{f:fig_tvd} strongly resemble those in Fig.~\ref{f:fig3}, indicating a similar degree
of saturation, but at a generally lower flux density. The variability index of the
flare in Fig.~\ref{f:fig_tvd} is 2.1.
These results are consistent with the data shown in Fig.~\ref{f:deepclouds}, where the spectra also
show that the internally variable object also achieves a lower flux density for the same maser
depth. If a different observable parameter is chosen, the brightest specific intensity of any
ray, results from the internally variable object are more spectacular: this parameter reaches
19900 times the saturation intensity in the variable case, compared to 7000 in the internally
uniform case. This parameter is only practically measurable in VLBI observation, and it is
apparent from the images in Fig.~\ref{f:deepclouds} that a small number of very high
intensity rays are not enough to compensate, in flux density, for many low intensity
neighbours.

It is also interesting to shift the observer's viewpoint to the `typical' position in the variable
unsaturated inversion case. The result is a low-gain light-curve. The highest flux density achieved
is only 7.00$\times$10$^{-7}$ with a variability index of 1.64. This index is lower than in the optimum
case, and indicates a more saturated light curve. In the lower depth range of $\tau=5-10$, the
light-curve is almost twice as bright as that in Fig.~\ref{f:typical}, but unlike that case, it
already shows significant signs of saturation.
At this point, we end our discussion of sample light curves, and
defer consideration of some variability parameters extracted from a much wider
range of data to Section~\ref{s:stats}.

\section{Variable Background Results}
\label{s:resvbcknd}

\begin{figure}
  \includegraphics[bb=380 90 620 790, scale=0.58,angle=0]{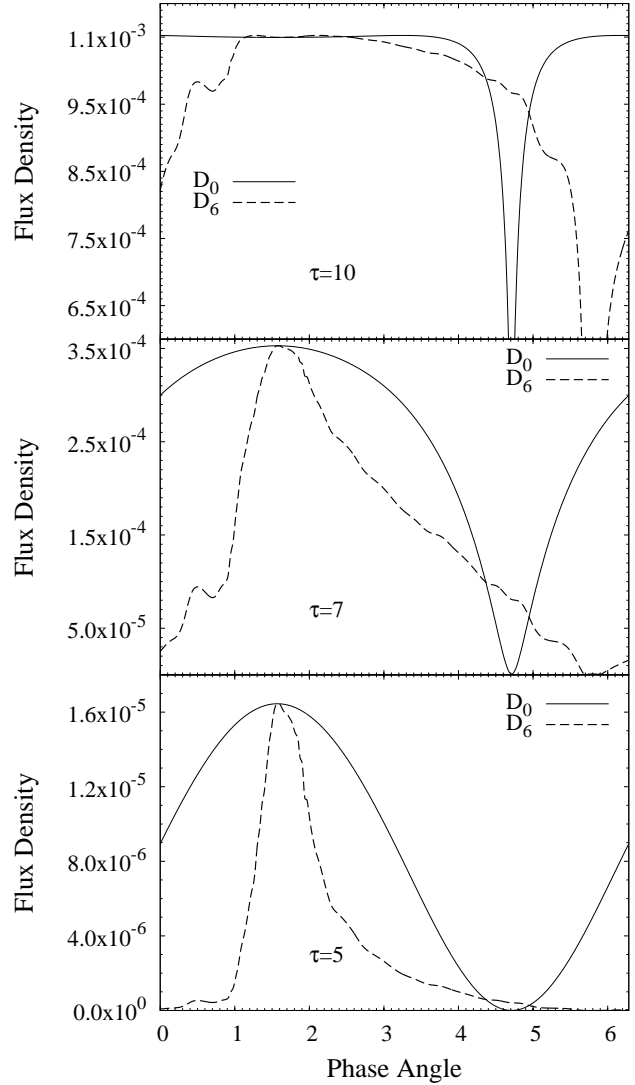}
  \caption{Maser light curves resulting from variation of the background radiation level that
follows the form of the driving functions $D_0$ and $D_6$. The minimum background level is
10$^{-9}$ and the range is 10$^{4}$. Panels represent calculations at different depth multipliers
of, from bottom to top, $\tau=$5.0,7.0 and 10.0.}
\label{f:bg_full}
\end{figure}

Here we show results for driving functions $D_0$ and $D_6$ fitted to eq.(\ref{eq:bgthing}). 
Function $D_6$ is chosen because it is particularly associated with variation of the
background radiation. Parameters
are now the minimum background level $i_{BG,min}$ and the range through which the background
intensitiy varies, $\Delta i_{BG}$. We begin with results for the prolate cloud with uniform inversion,
with the observer in the optimum viewing position. 
In Fig.~\ref{f:bg_full} we adopt the largest value of $\Delta i_{BG}$ covered by the
computations, that is
$\Delta i_{BG}=10^4$. This implies that $i_{BG,min}$ is 10$^{-9}$ of the saturation intensity, and the
maximum background attained is 10$^{-5}$. Each panel of Fig.~\ref{f:bg_full} shows light curves
for a different value of the depth multiplier, which increases from 5.0 in the lowest panel to
10.0 in the top panel.

In the lowest panel of Fig.~\ref{f:bg_full}, the cloud is only weakly saturated, but the light
curves here strongly resemble the original driving functions from Fig.~\ref{f:drivers}. This is
in contrast to the radiative pumping model, where the light curves would show strong near-exponential
growth, as in, for example, Fig.~\ref{f:typical}. As the depth rises, we see a trend that is qualitatively different to anything seen in the
work on radiative pumping. The trend is towards a constant saturation level (which rises somewhat
with depth) with a narrowing drop-out to a significantly lower flux density near the lowest
values of the driving function. The situation in the upper panel of Fig.~\ref{f:bg_full} could
perhaps be described as an `anti-flare', where the light curve has a duty cycle tending to 1, and
the high state of the curve is what an observer would mostly expect to see.

An expected trend with increasing saturation is a falling variability index, and this is
indeed what is seen in moving from the bottom to the top panel in Fig.~\ref{f:bg_full}, where
the respective variability indices are 8300, 421 and 1.66. If the depth multiplier is increased to
15.0, the maximum flux density rises to 2.12$\times$10$^{-3}$ and the variability index 
reduces to just 1.23. In fact, the duty cycle behaviour inverts from large ($\lesssim 1$) to
very small ($<0.1$) above a depth multiplier of $\sim 12$. This regime is discussed in
Section~\ref{s:stats}, but is not very interesting from the point of view of major flares because of
the very small values of the variability index.

\begin{figure}
  \includegraphics[bb=380 90 620 790, scale=0.58,angle=0]{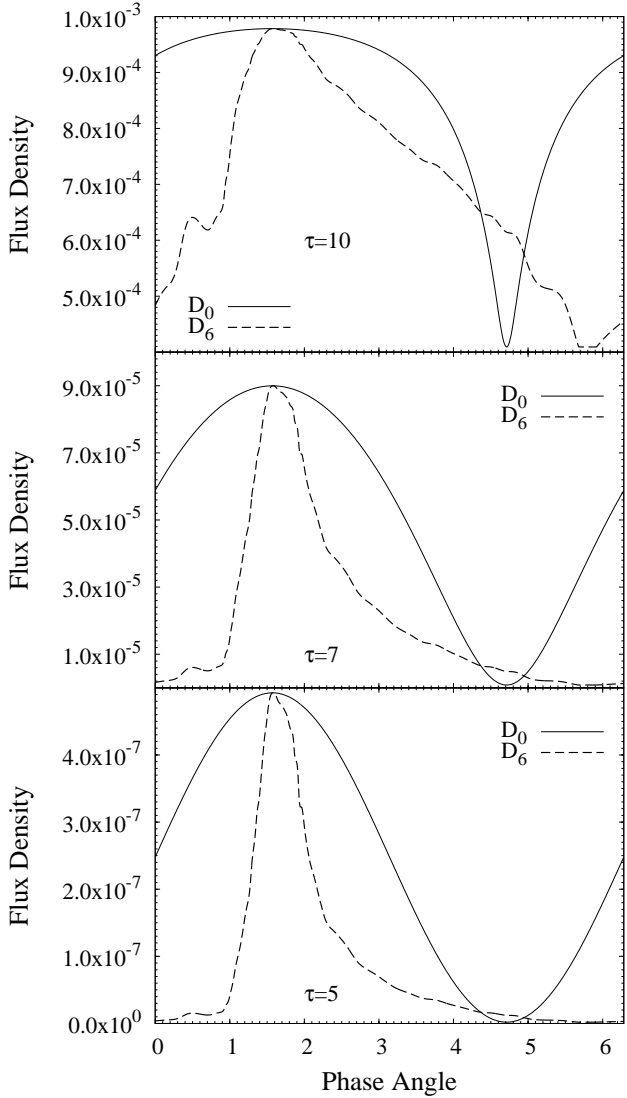}
  \caption{As for Fig.~\ref{f:bg_full}, but with a minimum background level of
10$^{-8}$ and a range of 250.}
\label{f:bg_small}
\end{figure}

As the change in background that has been applied so far might be considered rather large, further
light curves were computed with $i_{BG,min}=10^{-8}$ and $\Delta i_{BG}=250$. These additional light curves
are displayed in Fig.~\ref{f:bg_small}.
The results in Fig.~\ref{f:bg_small} are in line with expectations, given the high duty cycle effect
carried over from Fig.~\ref{f:bg_full}: saturation levels are lower, and the unsaturated behaviour,
where the maser light curves resemble the driving functions, is maintained to higher values of $\tau$.

Other variations do not change the qualitative behaviour shown in Fig.~\ref{f:bg_full} and
Fig.~\ref{f:bg_small}. Details follow the expectations from Sections~\ref{ss:viewpoint}-\ref{ss:tvdcloud}.
Viewpoints away from the cloud long axis lead to optically thinner behaviour and lower flux densities.
Replacement of the prolate cloud with oblate and spherical shapes yield lower flux densities generally, and
lower limiting flux densities at saturation. Use of clouds with internal variation in the unsaturated
inversion shift the maser response at a given value of the optical depth multiplier towards somewhat
thinner behaviour.

\section{Viewpoint Statistics}
\label{s:stats}

\begin{figure}
  \includegraphics[bb=45 35 620 225, scale=0.75,angle=0]{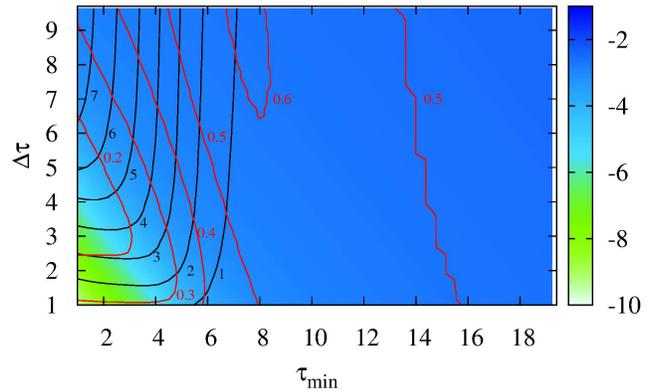}
  \caption{Three observable statisitcs as a function of minimum depth parameter and depth
parameter change. These are: the maximum flux density achieved over the period, where the colour
table represents the $\log_{10}$ of the flux density, the variability index (black contours with
levels indicating the $\log_{10}$ of this quantity), and the duty cycle (red contours with values
as marked). The background level is $i_{BG}=$10$^{-6}$, and results are for driving function
$D_0$ only, as applied to an internally uniform $\Gamma =0.6$ prolate cloud.}
\label{f:3parms_p_uni}
\end{figure}

\begin{figure}
  \includegraphics[bb=45 35 620 247, scale=0.75,angle=0]{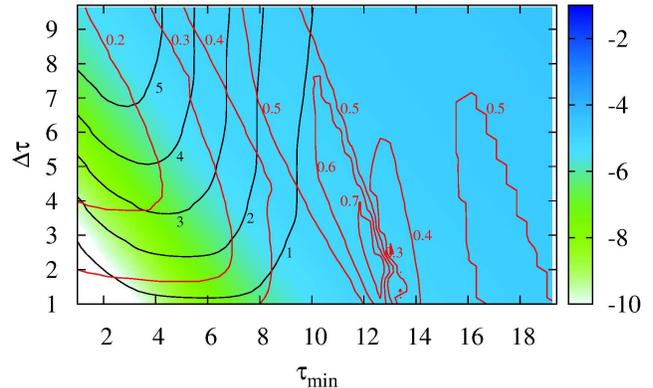}
  \caption{As for Fig.~\ref{f:3parms_p_uni}, but for the oblate cloud with internal
structure. The observer's viewpoint is in the $xy$ plane.}
\label{f:3parms_o_tvd}
\end{figure}

The light curves discussed in Section~\ref{s:reslightcurve} are just examples drawn from a wide
range of possible parameter values far too extensive to plot out in a paper of this type. Therefore,
we summarise a considerable quantity of data by extracting three important statistics from
the light curves: the varability index, the duty cycle (as defined
in Section~\ref{s:reslightcurve}), and the maximum flux density achieved. Each of these are
plotted as a function of both $\Delta \tau$ and $\tau_{min}$ for two extreme example types: a prolate cloud
with uniform unsaturated inversion, and an oblate cloud with the unsaturated inversion obeying a $1/r^2$
function. Here, both of these cloud styles are viewed
from close to their respective optimum observer's position. The maximum flux density achieved is shown on a
logarithmic scale via the colour palette in Fig.~\ref{f:3parms_p_uni}-\ref{f:3parms_o_tvd}. In
these same figures, black contours show the variability index and red contours, the duty cycle.

Results for the prolate cloud with uniform internal physical conditions are shown in
Fig.~\ref{f:3parms_p_uni}.
The first point to note is that saturating flux densities are achieved across most of the
plotted plane, so the great majority of masers on it would be observable, given the cloud
shape and privliged observer's position. The second point is that there is a minimum depth
multiplier - approximately 7.5 - above which it is impossible to generate a large amplitude
flare (variability index $>$10) via variation in the maser pumping. A third point is
that variability index contours
have an almost horizontal section that loops back to the $\Delta \tau$ axis, so there is also
a minimum variation to the depth multiplier required to give a certain flare amplitude: a
variability index of a million can be achieved with $\Delta \tau \geq 5$ for example. A fourth
point arises with
regard to the duty cycle: pumping variation is most unlikely to produce values much above
0.5: large $\Delta \tau$ and $\tau_{min}\sim 8$ are required to achieve 0.6, whilst duty
cycle values $<$0.3 are associated with unsaturated amplification.

As discussed above, the main effects of moving to an oblate cloud, even when viewed close
to its optimum plane, and of changing to an internally structured object, are to reduce
the maximum flux density of the maser and the variability index of flares. One more example is
shown in Fig.~\ref{f:3parms_o_tvd}: the counterpart of Fig.~\ref{f:3parms_p_uni} for an oblate cloud with a $1/r^2$
function in its unsaturated inversion.
The maximum variability index is now reduced to a number in the hundreds of thousands, and
the model has thinner behaviour generally as expected. However, lower black contours
representing variability indices of 10-100 survive to larger values of $\tau_{min}$
than in Fig.~\ref{f:3parms_p_uni}. There is also some interesting behaviour
in the duty cycle, where there is a steep gradient at low $\Delta \tau$: the value of
this parameter falls from above 0.7 to $\sim$0.3 over a small range of $\tau_{min}$ from 
approximately 12.0 to 13.0.

\begin{figure}
  \includegraphics[bb=58 70 620 260, scale=0.70,angle=0]{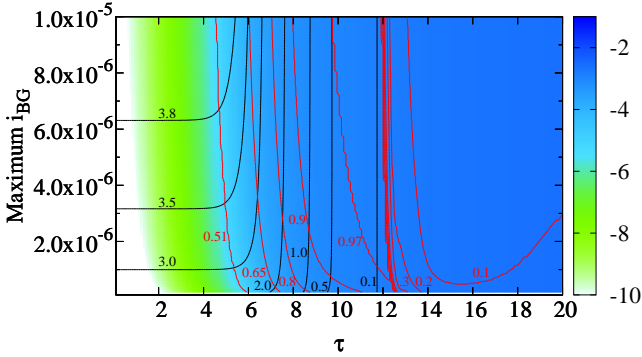}
  \caption{Statistics and cloud 
structure as for Fig.~\ref{f:3parms_p_uni}, but now plotted against optical
depth multiplier, $\tau$, and the maximum background intensity achieved by the driving
function in each model. The minimum background intensity in all models used by this figure
was $10^{-9}$ times the saturation intensity. The broad red region in this figure is a
zone of closely-spaced contours, where the duty cycle switches rapidly from very high
values to very low. This region is discussed in the text. The light curves plotted in
Fig.~\ref{f:bg_full} correspond to points in the present figure at the tops of vertical
slices with $\tau = 5,7$ and $10$.}
\label{f:bg3}
\end{figure}

When the background intensity is varied with time, rather than the pumping radiation, a rather
different pattern emerges from those in Fig.~\ref{f:3parms_p_uni} and Fig.~\ref{f:3parms_o_tvd}.
Although results are somewhat dependent also on the minimum intensity of background radiation
chosen, we show in Fig.~\ref{f:bg3} the same three statistics as in the previous two figures, now
as a function of optical depth multiplier ($x$ axis) and the maximum background intensity
used in each model ($y$ axis). The model cloud is the internally uniform prolate type, as used
in Fig.~\ref{f:3parms_p_uni}.
The $y$ axis values are scaled to the saturation intensity, and
the minimum background intensity used in all the models used to build Fig.~\ref{f:bg3} was
$1.0 \times 10^{-9}$.
The variability index of the driving functions is $10^{-7}/10^{-9}=100$ at the bottom of
Fig.~\ref{f:bg3} and $10000$ at the top. In fact, it appears that the variability
index of the maser response cannot exceed that of the driving function, unlike the variable
pumping case, but it can be reduced by saturation. 

Figure~\ref{f:bg3} can be conveniently divided into three regions from left to right along
the $\tau$ axis. The unsaturated models in the left-most part, where $\tau<4$, have a variability index
that varies with the maximum background intensity, but is almost independent of 
$\tau$, whilst the duty cycle is everywhere very close to 0.5. 
A particularly interesting region of
Fig.~\ref{f:bg3} exists for intermediate depth multipliers, in the approximate range 
4-14, where both the variability index and duty cycle become quite strong functions of
the optical depth multiplier, but comparatively weak functions of the maximum background
intensity. In this region, the variability index decreases monotonically
through the value of $10^{0.1}\simeq 1.26$ at a depth multiplier of $\sim 12$. The behaviour
of the duty cycle in this depth range is first to increase to values close to 1.0 (the
`anti-flare' behaviour exemplified by the top panel of Fig.~\ref{f:bg_full}), and then to
very rapidly fall as the narrow `hole' in the light curve becomes a low amplitude, but
still narrow, spike. On the high-depth side of this boundary, it is the lowest background
intensity that gives rise to the highest maser flux density, contrary to the situation found
at lower values of $\tau$. Although low values of the duty cycle are found in this central
portion, they are associated with such low levels of variability, that any flares of
signifcant variability index (a factor of 1.26 or more) would be associated with values
of the duty cycle above $0.5$.
In the high saturation region of Fig.~\ref{f:bg3} the variability index is low, and
continues to fall towards 1.0. The effect of the narrow spike in the light curve keeps
the duty cycle low, but, as the 0.1 red contour suggests, at extreme saturation the
duty cycle rises again, but no limiting value had been reached by a depth of $\tau=30$.

\begin{figure}
  \includegraphics[bb=320 100 560 570, scale=0.45,angle=0]{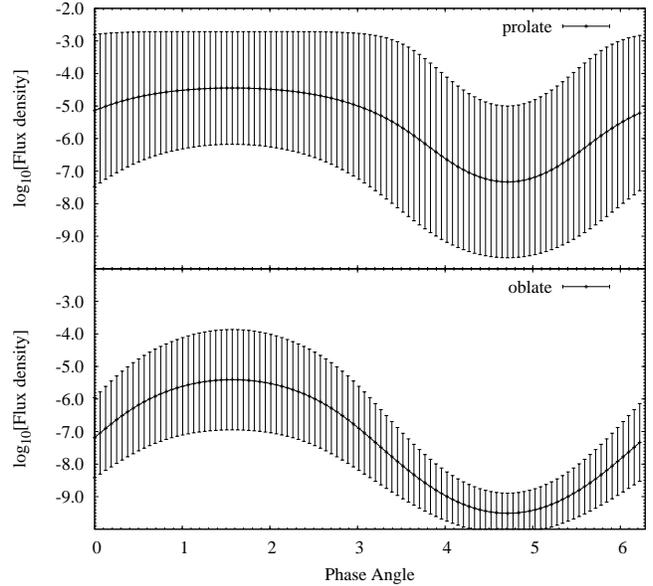}
  \caption{The base 10 logarithm of the mean light curve for $\tau_{min}=5$ and
$\Delta \tau=5$ for a prolate cloud with uniform unsaturated inversion (upper panel) and
an oblate cloud with $1/r^2$ unsaturated inversion (lower panel). Error bars are plotted in
both panels, marking the sample standard deviation, and the mean in both cases is over
1000 random observer's positions. Curves are responses to driving function $D_0$ only.}
\label{f:fig_stats}
\end{figure}

The final plot in this Section, Fig.~\ref{f:fig_stats} shows the effect of varying the observer's
viewpoint in a more general sense than in Section~\ref{ss:viewpoint}.
Formal solutions were obtained for 1000 randomly chosen observer's viewpoints for the
cases of a prolate cloud with uniform unsaturated inversions and an oblate cloud with the
unsaturated inversions obeying a $1/r^2$ function. The cloud shapes and structures therefore
follow those used in Fig.~\ref{f:3parms_p_uni} and Fig.~\ref{f:3parms_o_tvd}. The values
of $\tau_{min}$ and $\Delta \tau$ used were both 5, so the range in depth is the same
as in Fig.~\ref{f:fig4}. The curves plotted in Fig.~\ref{f:fig_stats} show the maser flux
density on a logarithmic scale to make the pattern of error bars clear; these show the
sample standard deviation and represent the envelope of likely values of the flux density.
Again, for reasons of clarity, only the response to the default driving function $D_0$ is
shown.

The maser response in the upper panel of Fig.~\ref{f:fig_stats} is clearly strongly influenced
by saturation, with the upper envelope limited to values of the flux density a little higher
than 10$^{-3}$.
The mean is less affected, but at the peak of the cycle it is $\sim$100 times weaker than the
saturation limit. The oblate cloud has a generally thinner envelope spanned by the error bars,
and is less affected by saturation. The latter observation is expected from the behaviour
discussed in Section~\ref{ss:cloudshape} and Section~\ref{ss:tvdcloud}. A typical cloud of this
type will yield a flare of greater variability index than the prolate cloud in the upper panel
of Fig.~\ref{f:fig_stats}, but will have a peak flux density about an order of magnitude
lower, a crucial consideration for detectability.

\section{Computing Details}
\label{timings}

The non-linear Orthomin($K$) algorithm (see \citetalias{2019MNRAS.486.4216G} and references therein)
is still the core routine used in the present work to calculate nodal
solutions. This algorithm has now been tested up to optical depth multipliers
of 50.0: almost certainly considerably beyond the validity of the CVR
approximation used in the present work. A single iteration job on the
DiRAC dial machine at Leicester takes 10\,s wall-time on the standard
(dirac25x) queue. As in previous work in Papers~1 and 2, time taken for
other computational tasks is negligible.

\section{Discussion}
\label{discuss}

Saturation has a profound smoothing effect on maser variability driven by both the variable
pumping and variable background mechanisms. This smoothing is stronger than in the case of
rotation. For example, Fig.~11 of Paper 2 shows that both prolate and oblate clouds with
$|\Gamma|=0.6$ can yield flares with a variability index $>$30 at a depth multiplier of
13.0. In the present work, Fig.~\ref{f:3parms_p_uni} and Fig.~\ref{f:3parms_o_tvd} indicate
that the variability index is $<$10 at this depth. For the specific example of Fig.~\ref{f:fig3}
that includes 13.0 in its depth range, the variability index is only 1.8.
Even strongly peaked driving functions, such as $D_1$ give rise to maser light curves
that are flatter near the peak when the depth is varied between 10.0 and 15.0. Their
troughs are less affected.

Modest saturation acts to increase the duty cycle function in
both the variable pumping and variable background mechanisms and, more generally,
to remove the special identity of individual driving functions in parts of
light curves that are at high flux density. Saturation
cannot, however, extend
the decay with respect to the rise time, or vice-versa.
Extreme saturation, with variable pumping, produces response curves that have a shape that strongly resembles the
driving functions: this is probably as expected, since a highly saturated maser is said to
behave like a linear converter of inversion to photons, for example, \citet{elitzurbook}.
The variability index of the flare continues to reduce with depth multiplier, and
the variability index for the 24.0-29.0 model (see Fig.~\ref{f:fig5}) is 
only about 1.22, barely above the 1.21 for the
driver. It is therefore difficult to get a large-amplitude flare from IR pumping in a saturated
maser cloud. The best possibilities for easing the restrictions imposed by saturation are
changing the internal distribution of the unsaturated inversion (or of the pumping rate)
and changing the saturation intensity of the maser. The latter effect does not change the
results presented in this work, which are scaled to this parameter, but would increase the
achievable dimensioned flux densities, as discussed in Appendix~\ref{a:fdunits}.

If the light curve of the IR pumping radiation is known at, or close to, the frequency of an
important pumping transition, then comparison with the light curve of the maser offers a
method of measuring the level of saturation in the maser cloud. Both the variability index
and the duty cycle depend on the base level of maser depth and the range over which it
is changed by the IR driving function. Therefore, these quantities can be read off a more
detailed version of a plot like Fig.~\ref{f:3parms_p_uni} by finding the best fit to the
black and red contours.

Shifting the observer's position away from privileged viewpoints close to the long axes of
clouds has a strong thinning effect. Whilst this allows for perhaps greater values of the
variability index, and lower duty cycles, it also reduces the mean flux density considerably: a factor of 4500
in the `typical position' example from Fig.~\ref{f:typical}, but two orders of magnitude
more generally from the averaged light curve for the prolate uniform cloud (upper panel
of Fig.~\ref{f:fig_stats}). These reductions suggest that the objects we see as
bright flaring masers probably are those that are oriented rather favourably towards us, and
another `alien' observer in a different part of the Galaxy might see a different subset
of objects. This suggestion is supported by the very poor performance of the pseudo-spherical
cloud in the shape comparison in Fig.~\ref{f:tricycle}. Changing from a prolate to an oblate 
cloud has effects that are quite consistent with those
found in \citetalias{2019MNRAS.486.4216G}. The oblate cloud gives somewhat less amplification and saturation from a
privileged viewpoint, but this is compensated for by a smaller standard deviation, so that
the loss of flux density from a random viewpoint is less pronounced.

In terms of a combination of raw flux density and high variability index,
the best maser flare candidates are obtained from
prolate clouds viewed near the optimum orientation. A variability index of over 10$^7$ can
be obtained (see Fig.~\ref{f:3parms_p_uni}) by starting from an object with extremely low
saturation ($\tau_{min}<1$) and applying a driving function with an amplitude of at least 7.
Such an object would still achieve a peak flux density close to the saturating level, and so
be detectable.

Use of clouds with an internal variation of the unsaturated inversion that behaves like
$1/r^2$ has a strong thinning effect: for a given depth multiplier, the internally variable
cloud gives a lower output, at least in terms of flux density. If instead the peak specific
intensity found in an image is chosen as the parameter to view, it is typically larger in
the internally variable cloud, at least from the privileged observer's viewpoint. However,
as the images in Fig.~\ref{f:deepclouds} suggest, the very high brightness rays are of very
limited number and area in the image of the internally variable cloud, and are not adequate to
make up for a larger number of somewhat less bright rays when computing the flux density.
The overall thinning effect is also supported by the histogram of nodes in various bands of
saturation in Fig.~\ref{f:deepclouds}. The cloud in Fig.~\ref{f:deepclouds}
is oblate, and viewed in the xy plane. A more general conclusion about the flux density
behaviour for an internally variable object of this type comes from the lower panel
of Fig.~\ref{f:fig_stats}. The typical level of the mean in this figure is an order of
magnitude below the uniform prolate object, so internal variation of this type is unlikely
to produce the most powerful flares. However, a form of internal variation that weights
unsaturated inversions towards the cloud surface might produce larger flux densities for
a given driving amplitude.

Variation of the background intensity is qualitatively different in at least two respects: firstly,
the variability index is always reduced below that of the background variation itself, particularly
at higher values of the depth multiplier. Observationally, maser variability must be accompanied
by radio continuum variability of at least the same amplitude at the frequency of the maser transition.
Secondly, the light curves that result from background variability have a characteristically higher
duty cycle at moderate maser depths, where the
variability index is $\gtrsim 1.26$. This effect is very apparent 
in Fig.~\ref{f:bg3}, and in the upper two panels of
Fig.~\ref{f:bg_full} and the top panel of Fig.~\ref{f:bg_small}.
The duty cycle parameter is likely a good discriminator between maser variability driven by
background variation, where values in the range 0.6-0.9 are common, and pumping-driven variability,
where values of the duty cycle $>$0.6 are rare. It is true that lower duty cycles can be obtained
at both lower, and higher, values of the maser depth multiplier, but in the former case detectability
is limited by lower, unsaturated, flux densities, whilst in the latter regime the variability index
is too low to be relevant to strong flaring events.

If a particular maser flare event has been positively identified as resulting
from background variability, then the duty cycle is potentially
also a very good diagnostic of saturation and/or the applied $\Delta i_{BG}$, provided that
it can be compared with the driving function, another need for high-cadence data at radio as
well as IR wavelengths. For aperiodic and temporally isolated flares, however generated, a modified duty cycle
parameter may still be defined. For example, if a flare is detected after a long quiescent period,
the modified parameter could be defined as the fraction of the total time for which the
light curve is above, say, the $5\sigma$ level of any minor variability during the quiescent
phase.

As a test of the duty cycle as a discriminator between driving mechanisms, and the ability to derive
flare parameters from our model, we attempt an analysis of the G107.298+5.63 periodic flare source,
where both IR and maser light curves exist. From Fig.~3 of \citet{2018IAUS..336...37S}, the maser has a
duty cycle of about 5 days in 35, or 1/7$\simeq$0.143, whilst the driving light curve, our function $D_1$
has a duty cycle of 0.2, so the maser duty cycle is reduced by a factor of $0.143/0.2=0.72$ with
respect to the IR driver. Since our Figures~\ref{f:3parms_p_uni}-\ref{f:bg3} are based on function $D_0$,
rather than $D_1$, we suggest that the most reasonable comparison is to use the same reduction factor,
placing G107.298+5.63 on the $0.5\times 0.72=0.36$ duty cycle contour. The -7.4\,km\,s$^{-1}$ methanol maser
spectral feature in G107.298+5.63 has a variability index of 120 \citep{2016MNRAS.459L..56S}, and it turns out
that we can find positions of coincidence for this value and a duty cycle of 0.36 in both Fig.~\ref{f:3parms_p_uni}
and Fig.~\ref{f:3parms_o_tvd}, but not in Fig.~\ref{f:bg3}. We therefore reject the variable background
mechanism for the flaring 6.7-GHz methanol masers in G107.298+5.63. For Fig.~\ref{f:3parms_p_uni}, the
model parameters read off the axes are approximately ($\tau_{min}=5$,$\Delta \tau=3$); corresponding
parameters in Fig.~\ref{f:3parms_o_tvd} are (7.5,4.5). We proceed to check that the model flux densities at
these positions is consistent with an observed value of approximately 50\,Jy. The relevant model flux
densities are $f_\nu=4.5\times 10^{-4}$ at (5,3) in Fig.~\ref{f:3parms_p_uni}, and $f_\nu=5.1\times 10^{-6}$
at (7.5,4.5) in Fig.~\ref{f:3parms_o_tvd}. These numbers can be checked for consistency via eq.(\ref{eq:fd_jansky})
with a source distance of $d_{kpc}$ = 0.76 and $Z=0.79$\,Hz. For Fig.~\ref{f:3parms_p_uni}, a cloud size
of $R_{AU}=4.8$ is required; for Fig.~\ref{f:3parms_o_tvd} a larger size of 45\,AU is needed.
Both these sizes are consistent with the 30-80\,AU region containing some clouds of both (methanol and H$_2$O) species
\citep{2016MNRAS.459L..56S}. However, the fact that multiple clouds exist in a region of this size suggests that the
smaller estimate of 4.8\,AU from Fig.~\ref{f:3parms_p_uni} is the more likely.

A similar analysis could possibly be attempted for S255-NIRS3, using the light curve from
\citet{2019PASJ..tmp..129U} as a driving function. We comment briefly here, but leave a quantitative
analysis to a future publication. Data associated with the IR light curve show that the variability
index (3400) of the methanol maser component at 6.42\,km\,s$^{-1}$, the brightest, is much higher
than that of the K$_s$ band IR light curve (23). This would appear to rule out a variable background
mechanism, but the variability index, and indeed the light curve, for the continuum at 6.7\,GHz are
unknown. We do note from Fig.~2 of Caratti o Garatti, Stecklum et al.
\citeyearpar{2017NatPh..13..276C} that the flux
density of the burst spectrum rises more quickly towards longer wavelengths than the pre-burst
counterpart, leading to larger variability indices at longer wavelengths. However, it would
be most unwise to extrapolate this trend very far. Purely from the point of view of the shapes
of the maser response and driving light curves, assuming the radio continuum follows the
IR pattern with sufficient amplitude, a variable background mechanism is favoured. We note that our function $D_6$ is
a rough approximation to the light curve in \citet{2019PASJ..tmp..129U}, and that the shape of
the maser response in Fig.~3 of \citet{2019PASJ..tmp..129U}, with its increased duty cycle with
respect to the driver, resembles somewhat the dashed ($D_6$) light curves
in the middle panel of Fig.~\ref{f:bg_full} and the top panel of Fig.~\ref{f:bg_small}.

Future work, in addition to a detailed analysis of S255-NIRS3,
should include a fuller analysis of the effects of a pump-dependent saturation intensity,
as discussed in Appendix~\ref{a:fdunits}. The only parameter in eq.(\ref{eq:satf}) that is scaled to the
saturation intensity is $i_{BG}$. If this becomes a function of the mean intensity in the pumping radiation,
the effect is to shift the chosen nodal solution from one of the family at the original $i_{BG}$ to one
from the family at lower (higher) $i_{BG}$ as the pump mean intensity rises (falls). Overall, the effect
is to couple pumping and background variability, though the effects are likely to be small, given the
weakness of the maser response to background variation. Programmes are also underway to model flares
generated by line-of-sight overlap of clouds and by shock compression. It is also hoped that more, and
better, IR data for maser flare sources of all types will enable detailed analyses
to be carried out, based on specific driving light curves, as we have attempted in a 
limited way for G107.298+5.63.

\section{Conclusions}
\label{conclusion}

Maser flares of very high variability index ($>$10$^7$) can be generated by variation in the
pumping radiation, provided that the maser cloud is initially very optically thin in the maser
transition, and the change in maser depth over a cycle of the light curve includes a substantial
portion of unsaturated growth. Detectability, or maximum flux density achieved, is optimised
if the maser clouds are of prolate spheroidal shape, perhaps elongated towards an almost filamentary
structure, and viewed close to the long axis of the cloud. Compared to rotation of irregular
objects, variable pumping is a poor mechanism of generating flares from already saturated maser
clouds.

Flares generated from variation in the background radiation at the maser frequency commonly
have high duty cycles if detectable flux density and strong flaring
variability are also requirements. The variability index
of the flare is probably always smaller than that of the background variation. As with pumping
variability, high initial saturation leads to a smaller variability index flare (or `anti-flare',
given the high duty cycle).

Clouds with an internal structure that makes unsaturated inversions larger towards the centre
of the cloud result in lower flux densities than uniform clouds of the same depth multiplier.
However, the structured clouds typically have a higher peak specific intensity (observationally,
the brightest pixel in a VLBI image). Clouds of this type would therefore have narrower beam angles,
and yield smaller spot sizes in VLBI images than internally uniform clouds of the same shape.

Further work should include full analysis of the case where the saturation intensity is itself
a function of the pumping radiation, additional weighting functions for internal structure and
full molecular energy level schemes to replace the current phenomenological pumping model.

\section*{Acknowledgments}

MDG and SE acknowledge funding from the UK Science and Technology Facilities
Council (STFC) as part of the consolidated grant ST/P000649/1 to the Jodrell Bank
Centre for Astrophysics at the University of Manchester. MDG acknowledges
financial support from the National Astronomical Research Institute of Thailand (NARIT)
whilst on sabbatical at their HQ in Chiang Mai, Thailand.
This work was performed, in part, using the DiRAC Data Intensive service at Leicester, operated 
by the University of Leicester IT Services, which forms part of the STFC 
DiRAC HPC Facility (www.dirac.ac.uk). The equipment was funded by BEIS capital 
funding via STFC capital grants ST/K000373/1 and ST/R002363/1 and 
STFC DiRAC Operations grant ST/R001014/1. DiRAC is part of the National e-Infrastructure.
Data used in this work was generated under DiRAC award dp124.

\bibliographystyle{mn2e}
\bibliography{MDGray_MN19_II}

\appendix

\section[]{Flux Density Units}
\label{a:fdunits}

Formal solutions in the present model produce specific intensities in a number of rays
that are scaled to the saturation intensity, $I_{sat}$, of the maser. In the IR pumping models, the
background level is $i_{BG}= I_{BG}/I_{sat} = 10^{-6}$. A flux density is computed via a standard
definition at a distance sufficient to make the rays almost parallel: by default, this distance
is 1000 times the cloud scale (see below). Rays originate on a disc, slightly more remote from the observer than the
cloud, that is designed to be larger than the cloud in any
orientation.
The slightly reduced version of this disc that passes through the cloud centre,
has a radius $R$ that we will call the cloud scale. The flux density is then
\begin{equation}
f_\nu = \sum_{j=1}^J i_j cos \theta_j \delta \Omega_j ,
\label{eq:raw_fd}
\end{equation}
where a particular ray has specific intensity $i_j$ in saturation units, originates at
an angle $\theta_j$ off-axis, as seen by the observer, and has a solid angle weighting
of $\delta \Omega_j$, which is almost the same for all rays (see \citetalias{2018MNRAS.477.2628G}).

When the cloud optical depth multiplier is zero, the cloud is transparent, and
$i_j = i_{BG}$ for all rays. The background flux density, assuming negligible
emission beyond the cloud scale is therefore,
\begin{equation}
f_{BG} = \pi i_{BG} (1/1000)^2 ,
\label{eq:back_fd}
\end{equation}
For the standard $i_{BG}=10^{-6}$ used in
the IR pumping part of the paper, eq.(~\ref{eq:back_fd}) reduces to
$f_{BG} = \pi \times 10^{-12}$. A moderately saturating value of the flux density is then of order 10$^{-6}$-10$^{-5}$,
whilst values of $>$10$^{-3}$ indicate rather strong saturation.

A useful method of placing this scaled formula into something more familiar to observers is to
place the distances in AU, and shift the observer to a standard distance of 1\,kpc. For distances
much larger than $R=1$\,AU, the flux density is proportional to $1/d^2$, and
\begin{equation}
f_{kpc} = f_\nu (R_{AU}/206265 d_{kpc})^2 .
\label{eq:kpc_fnu}
\end{equation}
The remaining issue is that $f_\nu$ is scaled to the saturation intensity of the maser transition. This
obviously depends on the particular transition studied. As a common example, we use the 6.7\,GHz maser
transition of methanol. The saturation intensity for a general maser is taken from \citet{mybook} as
\begin{equation}
I_{sat} = (Z + A_{ij} + C_{ij} + C_{ji})/(B_{ij}+B_{ji}),
\label{eq:isatty}
\end{equation} 
where $A_{ij},B_{ij}$ are respectively the Einstein A and B coefficients for transfer across the maser
transition, upper level $i$ and lower level, $j$. The $C$ coefficients are first-order collisional
rate coefficients in the same transition, whilst $Z$ is the all process rate coefficient for transfer
from level $i$ to all other levels excluding $j$ and acts, for current purposes, like extra spontaneous
emission. A data file from the Leiden Atomic and 
Molecular Database \citep{2005A&A...432..369S}, containing data from the CDMS \citep{ENDRES201695}, shows
that $A_{ij}=$1.593$\times$10$^{-9}$\,Hz is negligible compared to other terms on the right-hand side
of eq.(\ref{eq:isatty}). This, and
relations between the Einstein coefficients enable us to reduce the saturation intensity
to 
\begin{equation}
I_{sat} = \frac{2h\nu^3}{c^2 (1+g_i/g_j)}\left[
       \frac{Z+C_{ij}+C_{ji}}
              {A_{ij}}
                                       \right],
\label{eq:satint}
\end{equation}
where $\nu$ is the maser frequency of 6.66852\,GHz and $g_i/g_j = 11/13$ is
the ratio of the statistical weights of the levels (Leiden data file).
Second-order rate coefficients
for the maser transition itself are also dwarfed by the collisional terms to other levels that form
part of $Z$, and so can also be ignored at any density of collision partner. The result is that
eq.(\ref{eq:kpc_fnu}) can be reduced to the numerical version,
\begin{equation}
F_{kpc} \simeq 3500 Z (R_{AU}/d_{kpc})^2 f_\nu \;\;\mathrm{Jy}.
\label{eq:fd_jansky}
\end{equation}
The remaining difficulty is the value of $Z$. Downward radiative transitions from the upper maser
level make this at least 1.5$\times$10$^{-4}$\,Hz. Upward radiative rates are harder to gauge because
they involve a radiation mean intensity. A typical grey-body dust formula without attenuation or
geometrical dilution yields a rate, within the data from the Leiden file, that is unlikely
to exceed 10$^{-3}$\,Hz at any temperature ($<$250\,K) used in the methanol models by
\citet{2005MNRAS.360..533C}. Again using data within the Leiden file, downward second-order
rate coefficients \citep{2010MNRAS.406...95R} from the upper maser level sum to 3.21$\times$10$^{-10}$\,cm$^3$\,s$^{-1}$.
Upward contributions are harder to calculate, but similar, suggesting that an overall approximation
to $Z$ is $\sim$6.5$\times$10$^{-10} n_{H_2}$\,Hz for an H$_2$ number density in cm$^{-3}$. From graphs
in \citet{2005MNRAS.360..533C}, 0.0065\,Hz might be typical and 0.065\,Hz rather high without
quenching the maser.

Values of $Z$ estimated above would yield a constant saturation intensity for a cloud with constant
number density of H$_2$ and constant temperature. Adding collisional transfer to levels above the
Leiden set is unlikely to make much difference owing to increasingly adverse Boltzmann factors. However,
radiative transfer to levels in excited torsional states of methanol are also not included in the
Leiden set, and are vital to pumping the 6.7-GHz maser. Einstein A-values for those transitions that
couple the first and second torsionally excited states to the upper maser level can be found in
\citet{1993MNRAS.264..769C}. There are six such transition with A-values $>$0.01\,Hz, the
strongest being 0.297\,Hz. This is not
the whole story, as these transitions are upward with respect to the upper maser level, with a
rate $B\bar{J}$ that depends on a radiation field. If the radiation comes from dust, the rate due
to these lines will be approximately the sum of terms like,
\begin{equation}
B\bar{J} = A (e^{h\nu/(kT_d)} - 1)^{-1} (\lambda / \lambda_0)^{-p} ,
\label{eq:bj}
\end{equation}
where $T_d$ is the dust temperature, $A$ is the A-value and the wavelength ratio is a typical
dust weighting term with $\lambda_0 \sim 80$\,\micron \,and $p=2$. The whole of eq.(\ref{eq:bj})
might also be multiplied by geometrical dilution and/or attenuation factors. A quick calculation
using the six transitions introduced above for $T_d=250K$ yields a value significantly
faster than the collisional rate discussed above, that is $ \sum B\bar{J} \simeq Z = 0.79$\,Hz.
With this value, eq.(\ref{eq:fd_jansky}) yields a dimensioned flux density for a 10\,AU cloud
at 1\,kpc with $f_\nu=10^{-3}$ of 280\,Jy. However, through its dependence on $\bar{J}$, the
saturation intensity itself now depends on the pump rate, allowing the dimensioned flux density
to increase by a factor of typically a few over the rising part of the flare.

\end{document}